\documentclass[fleqn,usenatbib]{mnras}

\usepackage{newtxtext,newtxmath}

\usepackage[T1]{fontenc}
\usepackage{amsmath}
\usepackage{mathrsfs}

\usepackage{subcaption}

\DeclareRobustCommand{\VAN}[3]{#2}
\let\VANthebibliography\thebibliography
\def\thebibliography{\DeclareRobustCommand{\VAN}[3]{##3}\VANthebibliography}

\usepackage{graphicx}	
\usepackage{amsmath}	

\usepackage{microtype} 

\usepackage{etoolbox, siunitx}
\robustify\bfseries

\usepackage{booktabs} 

\usepackage[normalem]{ulem} 

\DeclareSIUnit\au{AU}
\DeclareSIUnit\Rsun{R_\odot}
\DeclareSIUnit\Rjup{R_\text{Jup}}
\DeclareSIUnit\Msun{M_\odot}
\DeclareSIUnit\Mjup{M_\text{Jup}}
\DeclareSIUnit\gyr{Gyr}
\DeclareSIUnit\ppt{ppt}
\DeclareSIUnit\ppm{ppm}

\title[Small CBP Migration]{Running The Gauntlet -- Survival of Small Circumbinary Planets Migrating Through Destabilising Resonances}

\author[Martin \& Fitzmaurice]{
David V. Martin$^{1,2}$ \&
Evan Fitzmaurice$^{1}$
\\
$^{1}$Department of Astronomy, The Ohio State University, 4055 McPherson Laboratory, Columbus, OH 43210, USA\\
$^{2}$NASA Sagan Fellow\\
martin.4096@osu.edu\\
}

\date{Accepted at MNRAS 2022 January 6. Received 2022 January 5; in original form 2021 October 19}

\pubyear{2021}

\begin{document}

\label{firstpage}
\pagerange{\pageref{firstpage}--\pageref{lastpage}}
\maketitle

\begin{abstract}

All of the known circumbinary planets are large ($R_{\rm p}\geq3R_\oplus$). Whilst observational biases may account for this dearth of small planets, in this paper we propose a theoretical explanation. Most of the known planets are near the stability limit, interspersed between  potentially unstable  $5:1$, $6:1$ and $7:1$ mean motion resonances with the binary.  It is believed that these planets did not form in situ, but rather migrated from farther out in the disc, and hence passed through these resonances. Planets are expected to migrate at a speed proportional to their mass, and a slower rate makes resonant capture and subsequent ejection more likely. Therefore, whilst large planets may be able to successfully ``run the gauntlet'', small planets may be imperiled. This hypothesis is tested using N-body integrations of migration in a truncated and turbulent disc. We discover that surprisingly none of the known planets exist interior to a fully unstable resonance. We demonstrate that resonant ejection of migrating planets may occur in nature, and that it does indeed disproportionately affect small planets, but it requires a highly turbulent disc and its efficiency is likely too low to fully explain a dearth of $R_{\rm p}<3R_\oplus$ planets.

\end{abstract}




\section{INTRODUCTION}\label{sec:introduction}

Fourteen transiting circumbinary planets have been discovered  in twelve systems (\citealt{martin2018,welsh2018,kostov2020,kostov2020c} and Table~\ref{tab:known_planets}). Nine  systems host a planet close to the stability limit at $\sim2.5-3.5$ times the binary separation \citep{martin2014,li2016,quarles2018,martin2018}, roughly corresponding to where the protoplanetary disc would have been truncated  by the tidal force from the binary \citep{Artymowicz1994,miranda2015,ragusa2020}. It is believed that these planets did not form in situ, owing to a highly turbulent environment this close to the binary \citep{Paardekooper2012,Lines2014,Meschiari2014,pierens2020,Pierens2021}. The favoured scenario is that the planets formed ex situ, i.e. farther out in the protoplanetary disc, before migrating inwards. A natural explanation for their observed positions is that their migration is halted or ``parked'' by a steep density gradient at the inner edge of the truncated disc  \citep{Pierens2008,pierens2013,thun2018,Penzlin2021}.


\begin{table*}
\begin{center}
\begin{tabular}{c|ccccc|ccccc}
\hline
Name        & $M_{\rm A}$ & $M_{\rm B}$ & $q_{\rm bin}$ & $P_{\rm bin}$ & $e_{\rm bin}$ & $R_{\rm p}$ & $M_{\rm p}$ & $P_{\rm p}$ & $e_{\rm p}$ & Period Ratio \\
& ($M_\odot$) & ($M_\odot$) & ($M_{\rm B}/M_{\rm A}$) & (Days) & & ($R_\oplus$) & ($M_{\oplus}$) & (Days) & & ($P_{\rm p}/P_{\rm bin}$)  \\
\hline
Kepler-16   & 0.69         & 0.20            & 0.29           & 41.08 & 0.16 & 8.27 &104.84 & 228.78 & 0.0069 & 5.57        \\
Kepler-34   & 1.05         & 1.02            & 0.97           & 27.80 & 0.52 & 8.38 & 69.89 & 288.82 & 0.180 & 10.39       \\
Kepler-35   & 0.89         & 0.81            & 0.91           & 20.73 & 0.14 & 7.99 & 44.78 & 131.46  & 0.042  & 6.34       \\
Kepler-38   & 0.95         & 0.25            & 0.26           & 18.79 & 0.10 & 4.20 & $<122$ & 105.60 & $< 0.032$ & 5.62       \\
Kepler-47b  & 1.04         & 0.36            & 0.35           & 7.45 & 0.023 & 3.05 & $<25.77$ & 49.53 & 0.021 & 6.65         \\
Kepler-47d  & 1.04         & 0.36            & 0.35           & 7.45 & 0.023 & 7.04 & 19.02 & 187.35 & 0.024 & 25.15         \\
Kepler-47c  & 1.04         & 0.36            & 0.35           & 7.45 & 0.023 & 4.65 & 3.17 & 303.14 & 0.044 & 40.70          \\
Kepler-64   & 1.53         & 0.41            & 0.27           & 20.00 & 0.21 & 6.10 & 168.70 & 138.32 & 0.054 & 6.92       \\
Kepler-413  & 0.82         & 0.54            & 0.66           & 10.12 & 0.037 & 4.35 & 67.00 & 66.26 & 0.12 & 6.55        \\
Kepler-453  & 0.93         & 0.19            & 0.20           & 27.32 & 0.051 & 6.30 & $< 15.88$ & 240.50 & 0.038 & 8.80        \\
Kepler-1647 & 1.22         & 0.97            & 0.80           & 11.26 & 0.16 & 11.64 & 483.00 & 1107.59 & 0.058 & 98.38 \\
Kepler-1661 & 0.84         & 0.26            & 0.31           & 28.16 & 0.112 & 3.87 & 17.00 & 175.06 & 0.057 & 6.22\\
TOI-1338 & 1.04         & 0.30            & 0.29           & 14.61 & 0.156 & 6.90 & 30.20 & 95.174 & 0.0938 & 6.52\\
TIC 172900988 & 1.24         & 1.20            & 0.97           & 19.66 & 0.448 & 11.25 & $\approx 822 - 981$ & $\approx180-210$ & $\lesssim0.09$ & $\approx 9.2 - 10.7$\\
\hline
\end{tabular}
\caption{Parameters of the known 14 transiting circumbinary planets. All have radii larger than $3R_{\oplus}$. Planet masses are typically poorly constrained or compatible with zero. TIC 17290098 (from hereon 1729), by virtue of only a single pair of transits on one passing, has a poorly constrained period. Discovery papers: 16 \citep{doyle2011}; 34 \& 35 \citep{welsh2012}; 38 \citep{orosz2012}; 47 \citep{orosz2012b,orosz2019}; 64 \citep{schwamb2013,kostov2013}; 413 \citep{kostov2014}; 453 \citep{welsh2015}; 1647 \citep{kostov2016}; 1661; \citep{socia2020}; 1338 \citep{kostov2020b}; 1729 \citep{kostov2020c}.}
\label{tab:known_planets}
\end{center}
\end{table*}

\begin{figure*}

    \includegraphics[width=\textwidth]{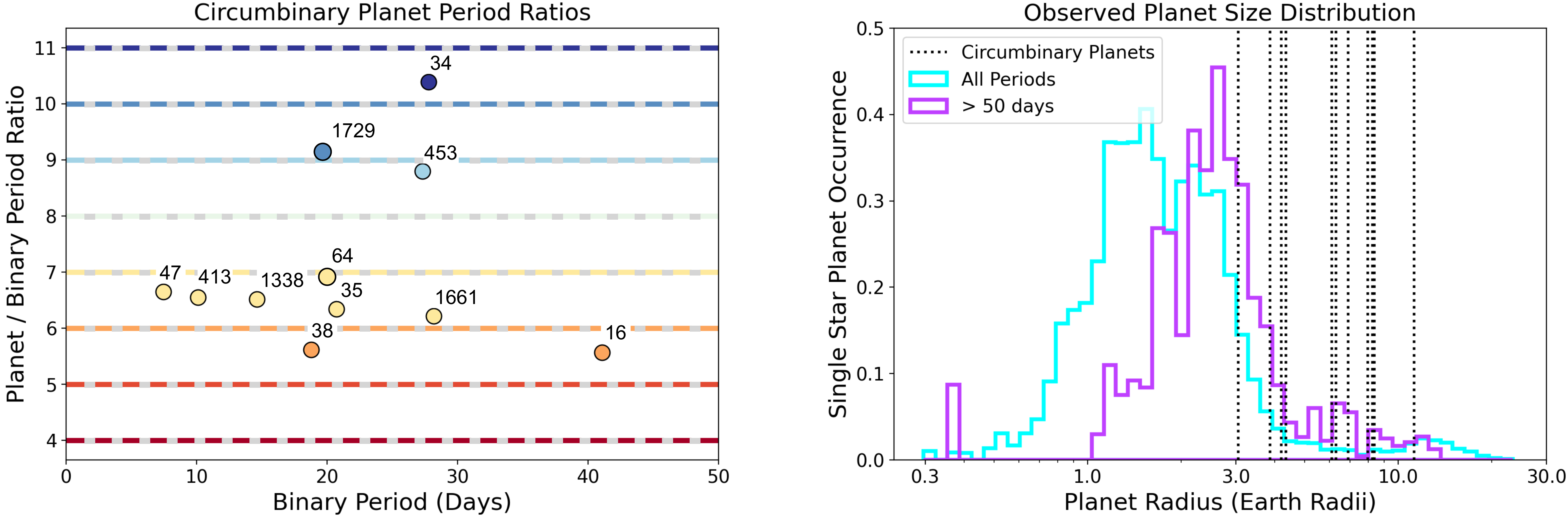}
    \caption{{\bf Left:} period ratios of the published circumbinary systems. For the three-planet system we only include the innermost planet (b). We exclude Kepler-1647, for which the period ratio is an outlier at 98.4. The planets are all located between $N:1$ mean motion resonances with the binary. The colours match those used later in Fig.~\ref{fig:lowest_order_stable_resonance}, where the planet's colour indicates the lowest order resonance crossed during migration (i.e. their period ratio rounded up to the nearest integer). {\bf Right:} observed size distribution of transiting planets around single stars, taken from the Exoplanet Archive. Distribution is shown for all orbital periods (cyan) and $P_{\rm p}>50$ days (purple), the latter of which matches the circumbinary planet periods. For comparison, the circumbinary planet radii are demarcated with dashed vertical lines. All observed circumbinary planets have radii above $3R_{\oplus}$, corresponding to the tail of the single star planet distribution. The same planets are shown as on the left plot for consistency, but the outer Kepler-47 planets and Kepler-1647 are also larger than $3R_{\oplus}$.}
    \label{fig:known_planets}
\end{figure*}

The stability limit itself is not  solely a function of planet-binary separation, but it has a complex shape dictated by mean motion resonances (MMR) with the binary. Essentially, any planet with $P_{\rm p}:P_{\rm bin} \lesssim 4$ will be unstable, but at wider orbits the first-degree $N:1$ resonances may or may not be unstable depending primarily on the planet and binary eccentricites, with islands of stability in between \citep{dvorak1984,holman1999,mardling2013,quarles2018,Lam2018}. 
All of the known circumbinary planets exist on stable orbits off first-degree resonances (Fig.~\ref{fig:known_planets}, left). However, if the planets did indeed migrate as we suspect, they would have passed through potentially destabilising resonances. For example, Kepler-16 has $P_{\rm p}:P_{\rm bin}= 5.57$, meaning that it passed through the $6:1$ and $7:1$ resonances, and so on.

Capture into these resonances, and potential instability and ejection, is more likely if the planet is migrating slowly \citep{Ketchum2011,Batygin2015,Mustill2016}.  Small planets follow Type-I migration, where they do not open a gap in the disc, and in this regime the migration speed is linearly proportional to the planet mass\footnote{ As opposed to Type-II migration, which we do not consider in this paper, where a massive planet opens a gap in the disc and then roughly follows the viscous evolution of the disc, which typically results in slower migration.} \citep{tanaka2002,lubow2010}. This means that  whilst larger planets may be able to ``run the gauntlet''\footnote{``Running the gauntlet'' means to pass quickly through a dangerous situation.} through trecherous resonances, small planets may be  imperilled by their slow migration. Indeed, of the known circumbinary planets none are smaller than $3R_{\oplus}$ (Fig.~\ref{fig:known_planets}, right), despite such small planets being the most abundant known around single stars \citep{petigura2013,fulton2017}. For now, detection biases and limitations probably account for most if not all of this dearth \citep{armstrong2014,martin2018}. However, this will hopefully change soon, as more sensitive transit detection techniques have been developed \citep{windemuth2019b,kostov2020b,martin2021}, in addition to advances in radial velocity (\citealt{martin2019}; Standing et al. under rev.) and microlensing detection \citep{Luhn2016,Bennett2016}. In anticipation, we ask the question:  {\it do we even expect small circumbinary planets to  exist?}

If small circumbinary planets truly do not exist, then there are two broad explanations: 1; they cannot form, or 2; they do form but cannot survive. With respect to formation, the simulations of \citet{Barbosa2020,Childs2021a,Childs2021b} do produce terrestrial circumbinary planets, although the presence of a giant planet may be an inhibitor. With respect to migration, \citet{sutherland2019}  demonstrated that multi-planet circumbinary systems may have a complex interplay between planet-planet and binary-planet resonances, potentially leading to ejections. They also analytically determined critical migration rates for planets to pass through $N:1$ resonances, however it was not shown which resonances were indeed unstable, nor how  resonant ejection would sculpt the mass/size distribution of circumbinary planets. The hydrodynamical simulations of \citet{thun2018,Penzlin2021} included an investigation of mass-dependent migration,  down to $48M_\oplus$ and $16M_\oplus$, respectively. They showed that more massive planets typically migrated closer to the star, owing to a circularisation of the disc. However, whilst smaller planets were expected to be parked farther from the binary, the orbits were not sufficiently wider to explain a lack of detections.

In this paper we propose and test a mechanism for the ejection of small circumbinary planets through resonant interactions precipitated by their slow migration. First, we quantify the level of instability for a given $N:1$ resonance, and thus determine for the known systems which (if any) unstable resonances were crossed during the planet's migration (Sect.~\ref{sec:stability}). We then construct N-body simulations with prescriptions for migration in a turbulent and truncated circumbinary disc (Sect.~\ref{sec:migration}). We first apply this in some select demonstrations of the different types which may arise (Sect.~\ref{sec:demonstration}), before applying our work to the known circumbinary systems to determine if small planets could have survived around the same binaries (Sect.~\ref{sec:applications}). In Sect.~\ref{sec:discussion} we discuss both the caveats and implications of our work, before concluding in Sect.~\ref{sec:conclusion}.

\section{Circumbinary Stability And Resonance}\label{sec:stability}


\subsection{Mean motion resonance}\label{subsec:resonances}

Consider two orbits with an integer ratio of periods:
\begin{equation}
    \frac{P_{\rm o}}{P_{\rm i}} = \frac{n_{\rm i}}{n_{\rm o}} =  \frac{p + q}{p},
\end{equation}
where   `o' = outer, `i' = inner, $n=2\pi/P$ is the mean motion of an orbit and $p$ and $q$ are integers. We define $q$ as the order of the resonance and $p$ as the degree of the resonance. For example, a $2:1$ period ratio is a first-order, first-degree resonance; $3:2$ is first-order, second-degree and $3:1$ is second-order, first-degree. 

By definition, when two orbits are in resonance at least one of their so-called resonant arguments will librate around a fixed value. This means that the orbits may coherently exchange energy and angular momentum over a long time, and gravitational interactions will be amplified. Assuming coplanar orbits, the resonant argument is defined as

\begin{equation}
\label{eq:resonant_argument}
    \phi = j_1\lambda_{\rm o} + j_2\lambda_{\rm i} + j_3\varpi_{\rm o} + j_4\varpi_{\rm i},
\end{equation}
where $\lambda$ is the mean longitude, $\varpi$ is the longitude of periapse and the $j$ coefficients are integers that follow the d'Alembert relation,
\begin{equation}
    \label{eq:dlamebert}
    \sum_{i=1}^{4}j_i=0
\end{equation}
(e.g. \citealt{murray1999}). For a mean motion resonance $j_1=p+q$ and $j_2=-p$, and then $j_3$ and $j_4$ are chosen to satisfy Eq.~\ref{eq:dlamebert}. A given resonance will have $q+1$ resonant arguments. However, resonance only requires that at least one of the arguments librates, and which angle(s) this is will depend on the mass ratios and eccentricities.



For a given resonance there is a range of orbital period ratios within which the system will be in resonance (i.e. $\phi$ librates). This is defined by the so-called resonance width $\Delta \sigma$. \citet{mardling2013} derived an expression for the width of first-degree $N:1$ resonances,

\begin{multline}
\label{eq:resonance_width}
    \Delta \sigma_N = \frac{6\mathscr{H}_{22}^{1/2}}{(2\pi)^{1/4}}\left[\left(\frac{M_3}{M_{123}}\right) + N^{2/3}\left(\frac{M_{12}}{M_{123}}\right)^{2/3} \left(\frac{M_1M_2}{M_{12}^2}\right) \right]^{1/2} \\
    \times \left(\frac{e_{\rm i}^{1/2}}{e_{\rm o}} \right) \left(1 - \frac{13}{24}e_{\rm i}^2 \right)^{1/2}\left(1-e_{\rm o}^2 \right)^{3/8} N^{3/4}e^{-N\xi(e_{\rm o})/2},
\end{multline}
where $\xi(e_{\rm 0}) = \cosh^{-1}(1/e_{\rm o}) - (1-e_{\rm o})^{1/2}$ and $\mathscr{H}_{22}^{1/2} = 0.71$ is an empirical scaling factor and we use the common shortenings of $M_{12} = M_1 + M_2$ and $M_{123} = M_1 + M_2 + M_3$. 

For circumbinary planets the inner mass ratio ($M_2/M_1\sim 0.1 - 1$) is much larger than that for a one star, two planet system (i.e. where $M_1>>M_2,M_3$). This means that from Eq.~\ref{eq:resonance_width} we can deduce that for circumbinary planets $\Delta \sigma_N$ is much larger for a given $N$. This is why we will investigate instability at high order resonances that normally are not discussed in the context of single stars. Eq.~\ref{eq:resonance_width} also suggests that no resonances can exist around circular binaries, which is not true \citep{DUNCAN1989,Malhotra1998,sutherland2019}. \citet{mardling2013} also determined that the libration period of a system in an $N:1$ resonance will be related to the resonant width by

\begin{equation}
    \label{eq:libration_period}
    P_{\rm N} = 2P_{\rm o}\Delta\sigma_N.
\end{equation}

\subsection{Stability map}\label{subsec:stability}

\begin{figure}
\begin{subfigure}{.5\textwidth}
  \centering
  \includegraphics[width=.99\linewidth]{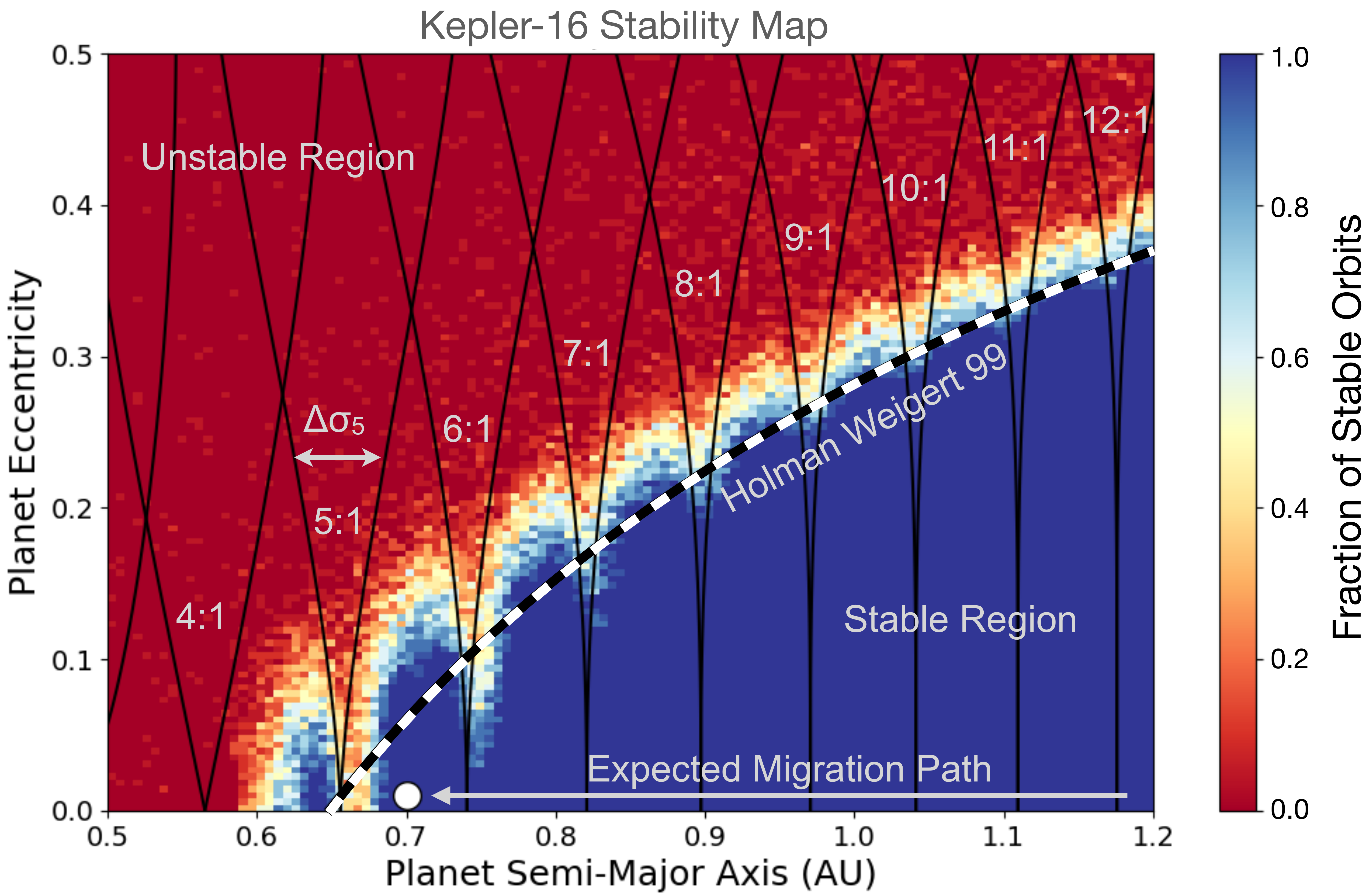}  
  \label{fig:sub-first}
\end{subfigure}
\caption{Stability map for Kepler-16. The colour is indicative of the fraction of stable simulations over a 10,000 year integration, where each simulation has a randomismed planet phase $\lambda_{\rm p}$ and argument of periapse $\omega_{\rm p}$. Over-plotted are the resonant widths from Eq.~\ref{eq:resonance_width}, derived by \citet{mardling2013}, where the width of each cone at a given planet eccentricity corresponds to $\Delta \sigma_N$, e.g. as illustrated for the $5:1$ resonance. We also overlay the classic \citet{holman1999} stability limit from Eq.~\ref{eq:holman99} with the \citet{quarles2018} planet eccentricity correction from Eq.~\ref{eq:ep_correction}. The $4:1$ resonance, and all interior orbits are unstable. The $5:1$ resonance is  ``fully unstable'', i.e. it is unstable even at $e_{\rm p}=0$, whereas the $6:1$ and higher-order resonances are  ``partially stable'' since they are stable for circular planets.}
\label{fig:stability_map}
\end{figure}

Planets orbiting too close to a binary will become unstable due to a process of resonance overlap \citep{Mudryk2006,Mardling2008}. For circumbinary planets the classic and often-cited stability limit of \citet{holman1999} is

\begin{equation}
\begin{split}
    \label{eq:holman99}
    a_{\rm crit}/a_{\rm bin} &= (1.60 \pm 0.04) + (5.10 \pm 0.05)e_{\rm bin} \\ &+ (-2.22 \pm 0.11)e_{\rm bin}^2  + (4.12 \pm 0.09)\mu_{\rm bin} \\ &+ (-4.27 \pm 0.17)e_{\rm bin}\mu_{\rm bin}  + (-5.09 \pm 0.11)\mu_{\rm bin}^2 \\ &+ (4.61 \pm 0.36)e_{\rm bin}^2\mu_{\rm bin}^2,
\end{split}
\end{equation}
were $\mu_{\rm bin} = M_{\rm B}/(M_{\rm A} + M_{\rm B})$. For a circular binary of equal mass stars, $a_{\rm crit} = 2.39a_{\rm bin}$.

The idea of a stability ``limit'' may be misleading though, as there is not in reality a sharp boundary dividing stable and unstable planets. A planet's stability is instead a more complex function of the orientation and phase of the orbits, and the planet's eccentricity.  These values also change over long secular timescales, for example apsidal precision \citep{farago10,cheng2019}. Importantly for this paper, a planet's stability is also a sensitive function of interactions with mean motion resonances with the binary. 

In Fig.~\ref{fig:stability_map} we create a stability map for Kepler-16. The binary parameters are held constant, and we vary $a_{\rm p}$ and $e_{\rm p}$ over a grid. At each point on this grid we run 20 N-body simulations over 10,000 years, where in each simulation the planet's starting phase $\lambda_{\rm p}$ and argument of periapse $\omega_{\rm p}$ are randomly drawn between 0 and 360$^{\circ}$, which is somewhat similar to the method of \citet{Lam2018}. The colour is indicative of the fraction of systems which remain stable ($e_{\rm p} < 1$) over the course of the simulation. 

Superimposed are the \citet{mardling2013} resonance widths from Eq.~\ref{eq:resonance_width}. These define the range of planet semi-major axes, centered on an exact commensurability, within which the planet will be in resonance. The resonant width increases with $e_{\rm p}$, which is why they have a cone-like structure in Fig.~\ref{fig:stability_map}. We also show the \citet{holman1999} stability limit from Eq.~\ref{eq:holman99}, which is roughly adapted to account for $e_{\rm p}$ based on \citet{quarles2018}:

\begin{equation}
    \label{eq:ep_correction}
    e_{\rm crit} = 0.8\left(1-\frac{a_{\rm crit}}{a_{\rm p}}\right).
\end{equation}

Our stability map is different to typical stability maps, e.g. Fig. 8 of \citet{quarles2018}, which are calculated with a single starting value of $\lambda_{\rm p}$ and $\omega_{\rm p}$. This is because there is not a sharp transition between the stable (blue) and unstable (red) regions of the parameter space; these angles are important dictators of stability\footnote{This effect was also investigated by \citet{quarles2018}, but just not in their stability map, for which fixed values were used.}. The \citet{holman1999} criterion (Eq.~\ref{eq:holman99}) roughly envelopes the stability region, but fails to encapsulate the finer details near period commensurabilities, and also any dependencies on the relative orbital and apsidal phases of the planet and binary. Resonances tend to ``increase'' instability, by which we mean they induce instability at lower planet eccentricities.

\subsection{Which resonances are fully unstable?}\label{subsec:fully_unstable_resonances}

\begin{figure}

    \includegraphics[width=0.49\textwidth]{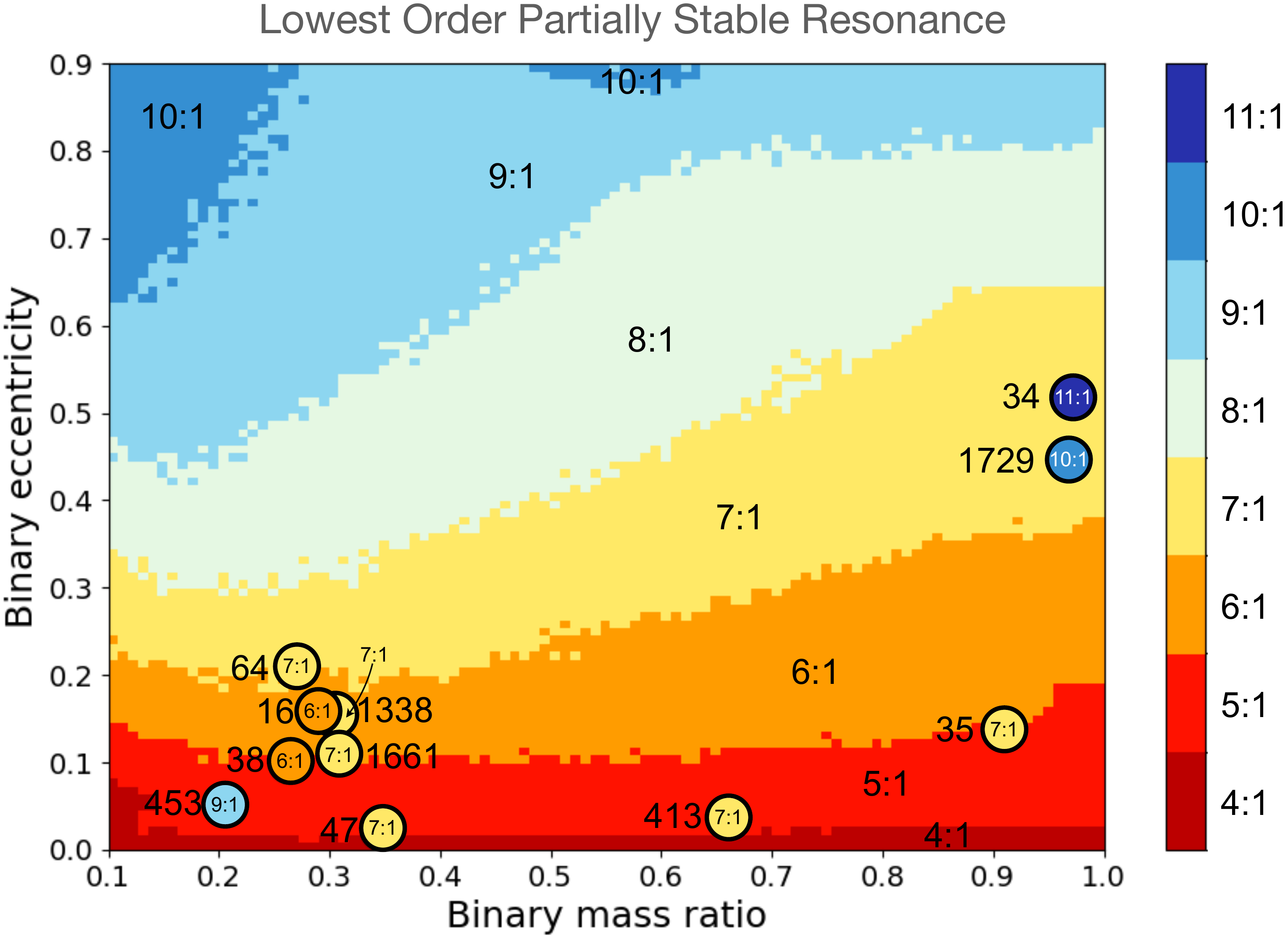}
    \caption{The lowest order partially stable resonance for a circumbinary planet around binaries of different mass ratios and eccentricities. A resonance is determined to be partially stable if an initially circular planet survives 1000 years of integration, for all period ratios between $N-0.1$ and $N+0.1$ times $P_{\rm bin}$ for a given resonance $N$, and for all 11 values of the starting phase $\lambda_{\rm p}$ between $0$ and $360^{\circ}$. The circles correspond to the known circumbinary planet hosts, labeled by their Kepler or TESS identifier. Inside each circle is the lowest order resonance which the planet would have migrated through, e.g. $6:1$ for Kepler-16 where $P_{\rm p}:P_{\rm bin}=5.57$.  The circle colour also matches the corresponding resonance. In all cases this resonance was partially stable. Otherwise put, {\it none of the known planets passed through a fully unstable resonance.}}
    \label{fig:lowest_order_stable_resonance}
\end{figure}


There has long been a general canonical notion that $N:1$ resonances for circumbinary planets are unstable. However, for this investigation of migrating circumbinary planets we need to know exactly {\it which} resonances are unstable, and to {\it what level}. We define a resonance as being ``fully unstable'' if even a circular planet becomes unstable at or near a given period commensurability. Alternatively, a resonance is defined as ``partially stable'' if at least circular planets may survive.

For the Kepler-16 example in Fig.~\ref{fig:stability_map} we see that the $4:1$ and $5:1$ resonances are fully unstable by this definition. Interestingly though the $6:1$ resonance is not fully unstable, as a sufficiently circular planet ($e_{\rm p} \lesssim 0.05$) may remain stable. Higher-order resonances are also not fully unstable, with the allowable planet eccentricity increasing with $N$.
The known planet Kepler-16b has a period ratio of $\sim5.57$. By our definition, it therefore did not migrate to this position through any fully unstable resonances. 

We now seek to determine, for any given binary, the lowest order partially stable resonance (e.g. $6:1$ for Kepler-16). This will be a function of the binary mass ratio $q_{\rm bin}=M_{\rm B}/M_{\rm A}$ and the binary eccentricity $e_{\rm bin}$. Due to scale invariance, there should not be a dependence on the absolute stellar masses.

We create a 20x20 grid of binary eccentricity and mass ratio. At each grid point we create a binary with a period of $P_{\rm bin}=20$ days and total mass $m_{\rm AB}=1.5M_\odot$. Then for each binary we loop through eight $N:1$ resonances from $N=4$ to $N=11$. At each resonance we test 21 planets with periods equally spaced between $(N-0.1)$ and $(N+0.1)$ times $P_{\rm bin}$, since the resonances may be slightly offset from exact commensurability \citep{mardling2013}. Then, for each planet period we test 10 starting phases between $0$ and $360^{\circ}$, since the relative planet-binary phase is important for stability \citep{quarles2018,Lam2018}. The planet is always initially circular, giving it the ``most chance'' to be stable. Each simulation is run for 10,000 years and the planet is deemed unstable if it is ejected ($e_{\rm p}>1$). In total $20\times20\times8\times21\times10 = 672,000$ simulations are run. 

We call a resonance fully unstable if any single one of these circular planets is ejected. We justify this in the context of our investigation of migrating planets; migration will pass a planet through all possible relative orbital phases with the binary, and so if any of them are unstable the planet will get ejected (if the migration is slow enough, see Sect.~\ref{sec:migration}). The results are plotted in Fig.~\ref{fig:lowest_order_stable_resonance}. The colour indicates the lowest order partially stable resonance. Unsurprisingly, the binary eccentricity is the biggest factor in determining resonant stability, as a higher $e_{\rm bin}$ pushes the stability limit farther out.

Over the top we plot the 11 binaries\footnote{For Kepler-47 we only consider the inner-most planet. We also ignore Kepler-1647b, since at a period ratio of nearly 100 this planet did not migrate to anywhere near the stability limit and disc inner edge. For TIC 12853201 we use the lowest possible period ratio from \citet{kostov2020c} of 9.2.} known to host planets near the stability limit. The colours of these circles correspond to the period ratio of the system rounded up to the nearest integer (e.g. $5.57 \rightarrow 6$ for Kepler-16) as this indicates the last resonance crossed by the planet before it stopped migrating. 

The interesting and perhaps surprising result we take from Fig.~\ref{fig:lowest_order_stable_resonance} is that none of the known planets exist on orbits interior to a fully unstable resonance. In three cases (Kepler-16, -38 and -64) the next interior resonance would be fully unstable, but they managed to park just before it. All other planets have a buffer of at least one resonance before the nearest fully unstable resonance.

This result suggests that none of the known planets had to ``run the gauntlet'' by migrating quickly through an unstable resonance. This would imply that their migration rate (and its dependence on the planet mass) is irrelevant in this context. The caveat to this argument is that the planetary eccentricity must remain sufficiently low through its migration, which is something we test in the subsequent section.




\section{Circumbinary planet migration}\label{sec:migration}

Planets in protoplanetary discs feel torques which will cause them to migrate over time, typically inwards \citep{Goldreich1979,Ward1986,lin1996,ward1997,Nelson2018}. This phenomena has been studied by many authors over the years, including for circumbinary planets \citep{Pierens2008,pierens2013,Kley2015,thun2018,Penzlin2021}.

Circumbinary planets form and migrate in a  disc which has been truncated by the tidal force of the two stars \citep{Artymowicz1994,miranda2015}. This truncation occurs at $\sim2.5-3.5\times a_{\rm bin}$, with a dependence on the binary eccentricity, mass ratio, disc properties  (e.g. density, viscosity, temperature) and can also have a dependence on the planet mass \citep{thun2018,Penzlin2021}. The conventional story, which has been demonstrated repeatedly by hydrodynamical simulations, is that the planets initially have an inwards migration due to a Lindblad torque, which approximately scales with the disc surface density $\Sigma$. When the planets get close to the disc edge  there will be a steep positive surface density gradient $d\Sigma/dr$. Since the co-orbital torque approximately scales with this gradient, it will counter-act the inwards Lindblad torque. This causes the planet to  be ``parked'' slightly interior to the peak of the disc density profile. The disc truncation (and hence parking location) has a similar scaling with $a_{\rm bin}$ as the N-body stability limit, and hence there is a natural explanation for the propensity of circumbinary planets to be found near the stability limit \citep{martin2014,li2016,martin2018}, although more observations are required to confirm the statistical significance of this trend.

Broadly speaking there are two types of migration, creatively named Type-I and Type-II. Type-I is for low mass planets which are embedded in the disk. Type-II is for high mass planets which carve a gap in the disc and migrate at a different (typically slower) rate. The transition between the two occurs roughly between a Saturn and Jupiter mass, although it will depend on the local disc properties. The known circumbinary planets, at least those close to the stability limit, are all smaller than $\approx 0.5M_{\rm Jup}$.  \citet{Pierens2008} argue that all of these planets underwent Type-I migration. Their hydrodynamical simulations show that if the planet carves its own gap in the disc (i.e. Type-II migration) this quenches the steep density gradient near the inner disc edge. This removes the parking mechanism since there will not be a strong outwards co-orbital torque, and hence the planet gets too close to the binary and is ejected (typically from the $4:1$ resonance). In this paper we are testing planet masses equal to or less than the known planet masses, and hence in within this mass range it is reasonable to assume that they underwent Type-I migration.

In Sect.~\ref{subsec:nbody_migration} we show how migration is implemented in an N-body code. Then in Sect.~\ref{subsec:torques} we show our prescriptions for Type-I migration torques. In Sect.~\ref{subsec:disc_model} we demonstrate our simple model of a truncated disc. Finally, in Sect.~\ref{subsec:stochastic_forcing} we describe the effect of stochastic forcing, used to model the effect of turbulence in the disc on the planet. These effects are then later demonstrated in Sect.~\ref{sec:demonstration}.

\subsection{N-body migration in \textsc{rebound}}\label{subsec:nbody_migration}

To understand the physics of migration people often use state of the art hydrodynamical simulations, e.g. \textsc{PLUTO} and \textsc{FARGO}. A benefit of such simulations is that it can capture the fundamental  interactions between the gas, dust, stars and planet. A disadvantage is that they are computationally expensive. Furthermore, regardless of how detailed the physical model is  (which itself will have underlying uncertainties), the outputted behaviour of the disc and planet will ultimately depend on the inputted parameters, which will inevitably have a large degree of uncertainty.

An alternate method, which we apply in this paper, is to use an N-body integrator with additional dissipative forces added to mimic migration at a certain rate. This has been done for planets around single stars \citep{rein2012FOURTHREE,rein2012PERIODRATIOS,bodman2014,Teyssandier2020,Huang2021,MacDonald2021,Huang2021} and binaries \citep{sutherland2019}. The rate of migration expected has been derived for example by \citet{ward1997,Papaloizou2000,tanaka2002,lee2002,Paardekooper2010,lubow2010,Paardekooper2011}, including calibrations against hydrodynamical simulations.  The main advantage of this N-body method is that it is significantly faster than hydrodynamical simulations. One may therefore run comprehensive suites of simulations over a large swathe of parameters.

In this paper we use the \textsc{rebound} N-body package \citep{rein2012REBOUND} to simulate the planet and binary orbits, using the WHFast integrator \citep{rein2015WHFAST} with a sufficiently short $0.001$ yr timestep to give results comparable to the more accurate (but significantly slower) IAS15 integrator \citep{rein2015IAS15}. We implement migration using the add-on package \textsc{reboundx} \citep{Tamayo2020}, specifically the \textsc{modify\_orbits\_forces} module. The planet's semi-major axis evolves from its starting value $a_{\rm p,0}$ over time $t$ according to

\begin{equation}
    \label{eq:a_p_function}
    a_{\rm p} = a_{\rm p,0}\exp\left(-\frac{t}{\tau_a}\right).
\end{equation}

In \textsc{reboundx} one sets $\tau_a$ as a parameter. In some studies, e.g. \citet{rein2012FOURTHREE,rein2012PERIODRATIOS}, $\tau_a$ is maintained at a constant value for a given simulation. We will instead follow  \citet{sutherland2019}, where $\tau_a$ is calculated at each timestep of the N-body integration based on predicted migration torques induced by the disc.

The migration rate is calculated according to

\begin{equation}
    \label{eq:tau_a}
    \tau_a = \frac{J_{\rm p}}{T_{\rm tot}},
\end{equation}
where $T_{\rm tot}$ is  the sum of all torques on the planet from the disc (calculated in Sect.~\ref{subsec:torques}) and 
\begin{equation}
    \label{eq:planet_angular_momentum}
    J_{\rm p} = M_{\rm p}\sqrt{a_{\rm p}G(M_{\rm A} + M_{\rm B} + M_{\rm p})}
\end{equation}
 is the planet's angular momentum. The disc will also act to damp the planet's eccentricity on a timescale of $\tau_e$. We invoke a commonly used simple relation of 

\begin{equation}
    \label{eq:tau_e}
    \tau_e = \frac{\tau_a}{K},
\end{equation}
where $K$ is a constant. In this paper we will use the typical value of $K=10$, as found by previous studies \citep{lee2002,Kley2004}, but we will test the effects of varying $K$ in Sect.~\ref{subsec:eccentricity_damping}.

\begin{figure}

    \includegraphics[width=0.49\textwidth]{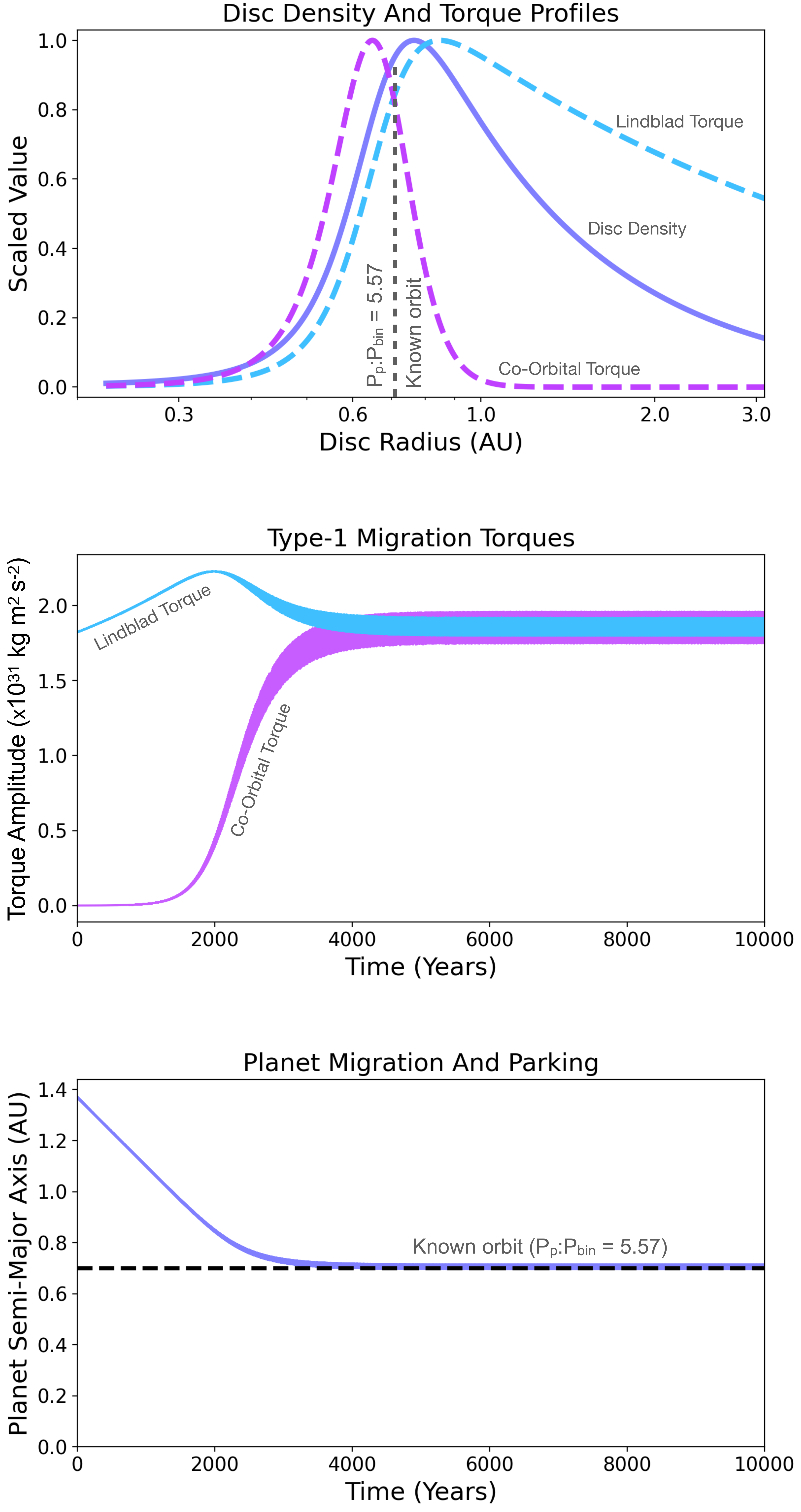}
    \caption{Demonstration of circumbinary planet migration and parking in Kepler-16. {\bf Top panel:} surface density profile of a truncated circumbinary disc (Eq.~\ref{eq:disc_profile}) and the expected magnitudes of the  Lindblad (Eq.~\ref{eq:lindblad}) and co-orbital torques (Eq.~\ref{eq:coorbital}). All have been scaled between 0 and 1 to demonstrate the functional dependence on $a_{\rm p}$. At $a_{\rm p}=0.7$ AU the two torques are equal in magnitude and opposite in sign, and hence the planet should park here. {\bf Middle panel:} N-body simulation of a planet starting near 1.4 AU and migrating inwards, showing the change in the Lindblad (blue) and co-orbital (purple) torques over time until they balance and the planet parks. {\bf Bottom panel:} time-variation of $a_{\rm p}$ until parking.}
    \label{fig:demo_simple}
\end{figure}

\subsection{Torque prescriptions}\label{subsec:torques}

In this paper we will only model Type-I migration, given the small planets we are focused on (see Sect.~\ref{subsec:discussion_caveats}). There are two primary forms of Type-I migration torques. First, the Lindblad torque comes from the gravitational back-reaction of spiral density waves launched by the planet at mean motion resonances with the disc. 

Second, the co-orbital torque is created by the disc material nearest to the planet, librating within the co-orbital ``horseshoe'' region. This second torque requires an asymmetry between the material interior and exterior to the planet, and hence is typically negligible except in the case of a steep disc density gradient.
 
 We use prescriptions from \citet{tanaka2002,lubow2010} for both the Lindblad and co-orbital torques. These torques are derived as a function of the planet's distance to the barycentre, which we approximate as the planet semi-major axis $a_{\rm p}$. The Lindblad torque is

\begin{equation}
    \label{eq:lindblad}
    T_{\rm L}(a_{\rm p}) = -\Sigma(a_{\rm p})\Omega_{\rm p}^2a_{\rm p}^4\left(\frac{M_{\rm p}}{m_{\rm AB}}\right)^2\left(\frac{a_{\rm p}}{H} \right)^2,
\end{equation}
where 
\begin{equation}
\label{eq:orbital_frequency}
    \Omega_{\rm p} = \sqrt{\frac{GM_{\rm AB}}{a_{\rm p}^3}}
\end{equation}
 is the Keplerian orbital frequency, $G$ is the gravitational constant, $\Sigma(a_{\rm p})$ is the disc surface density (as a function of radius, approximated by $a_{\rm p}$) and $H$ is the disc height \citep{lubow2010}. For typical minimum mass solar nebula (MMSN) the disc height profile is $H=ha_{\rm p}$, for some constant aspect ratio $h$. 

\citet{lubow2010} note that the simple assumptions made in deriving of Eq.~\ref{eq:lindblad} mean that whilst the migration rate is accurate, its direction is uncertain. We will always take it as being inwards (hence the negative sign), which is typical for Lindblad torques \citep{Ward1986,ward1997} and matches circumbinary hydrodynamical simulations such as \citet{pierens2013,thun2018}.

For the co-orbital torque we use 

\begin{equation}
    \label{eq:coorbital}
    T_{\rm co}(a_{\rm p}) = s\Sigma(a_{\rm p})\Omega_{\rm p}^2a_{\rm p}w^3\left(\frac{\Delta\Sigma}{\Sigma} - \frac{\Delta B}{B}\right),
\end{equation}
where $B$ is the Oort constant and $B=\Omega_{\rm p}/4$ for a Keplerian disc and $s$ is a scaling factor. The co-orbital torque is calculated over the so-called horseshoe region near the planet, which has width $w$. We approximate this width to be equal to one Hill radii; $w=a_{\rm p}[M_{\rm p}/(3m_{\rm AB})]^{1/2}$. We also define $\Delta\Sigma=\Sigma(a_{\rm p}+w) - \Sigma(a_{\rm p}-w)$ and $\Delta B=B(a_{\rm p}+w) - B(a_{\rm p}-w)$ as differentials across this horseshoe region. At the inner edge of a truncated circumbinary disc there is a steep positive $d\Sigma/dr$ gradient, for which the co-orbital torque will be maximised and in the outwards direction, opposing the Lindblad torque.

The scaling factor $s$ is something we have added with respect to the original derivation in \citet{lubow2010} and is calculated such that the maximum value of $T_{\rm co}$ (as a function of $a_{\rm p}$) is the same as the maximum of $T_{\rm L}$. For circumbinary planets to park there must be a value $a_{\rm p}$ for which the two torques are equal in magnitude (but opposite in sign). Whilst the two equations (Eq.~\ref{eq:lindblad} and \ref{eq:coorbital}) without $s$ will be on roughly the same order of magnitude, it is not guaranteed that such a balance will occur. We justify this scaling by reminding the reader that the purpose of this paper is not to {\it prove} that circumbinary planets park, as that both requires and was done successfully using hydrodynamical simulations (e.g. \citealt{Pierens2008,pierens2013,thun2018,pierens2020}). We instead want to create a simple model to replicate the fundamentals of the planet's behaviour as discovered by these authors.

These two torques are implemented in \textsc{rebound} simply by modifying the migration rate $\tau_{\rm a}$ through $T_{\rm tot} = T_{\rm L} + T_{\rm co}$ in Eq.~\ref{eq:tau_a}.

\subsection{Circumbinary disc model}\label{subsec:disc_model}

Circumbinary discs will be truncated by the tidal forces of the central binary, at a distance of roughly $2.5-3.5$ times the binary separation \citep{Artymowicz1994}. We use a truncated disc profile from  \citet{pierens2013}:

\begin{equation}
    \label{eq:disc_profile}
    \Sigma(r) = f_{\rm gap}\Sigma_0r^{-3/2},
\end{equation}
where $\Sigma_0$ is a constant which determines the total disc mass, $r^{-3/2}$ is a typical power law and $f_{\rm gap}$ is used to model the truncation of the disc via

\begin{equation}
    \label{eq:f_gap}
    f_{\rm gap} = \left(1 + \exp\left[-\frac{r-R_{\rm gap}}{0.1R_{\rm gap}}\right] \right)^{-1}. 
\end{equation}
The parameter $R_{\rm gap}$ determines how wide the inner gap is, and is typically set to $R_{\rm gap}=2.5a_{\rm bin}$ in their simulations.  Our disc profile is plotted in Fig.~\ref{fig:demo_simple}.  Note that Eq.~\ref{eq:disc_profile} is used as a starting condition in the hydrodynamical simulations of \citet{pierens2013} and similar studies. They then let the disc evolve under the influence of the binary. The result is that their disc profiles have the same rough shape as ours, i.e. a steep inner gradient followed by a power law decay, but are not as smooth. They also have finer details such as an eccentric cavity, whereas our symmetric profile effectively assumes a circular cavity. We discuss this caveat in Sect.~\ref{subsec:discussion_caveats}.

In Fig.~\ref{fig:demo_simple} we plot the predicted Lindblad and co-orbital torques as a function of the planet's position in the disc. From their respective equations, it is evident that the  Lindblad (Eq.~\ref{eq:lindblad}) torque is a function of the density whereas the co-orbital torque (Eq.~\ref{eq:coorbital}) is a function of the density gradient. We see in Fig.~\ref{fig:demo_simple} that the two torques balance at the inner disc edge, where there is a large density gradient. This   replicates the hydrodynamical simulations of \citep{Pierens2008,pierens2013,thun2018,Penzlin2021}.

\subsection{Stochastic forcing}\label{subsec:stochastic_forcing}

Near the inner edge of a truncated circumbinary disc the environment is expected to be highly turbulent, due to the tidal influence of the nearby binary. In fact, this is one of the main arguments against in situ planet formation \citep{Paardekooper2012,Lines2014,Meschiari2014,pierens2020,Pierens2021}. In our N-body simulations we use a simple model for disc turbulence known as stochastic forcing, following the method first developed by \citet{Rein2009,Rein2010PhD}. \citet{Adams2008,Ketchum2011} also developed independent methods for replicating the effects of turbulence for N-body simulations.

A first-order Markov process creates correlated, continuous noise that is applied to the planet as an additional force, mimicking turbulence. This is done independently to both $x$ and $y$ coordinates, with the force strength decaying exponentially with an auto-correlation time $\tau$. Simply put, a planet is constantly pushed in random directions within the plane of the disc, undergoing effectively a random walk. This randomness is in addition to the smooth migration from the Lindblad and co-orbital torques from Sect.~\ref{subsec:torques}.

The  acceleration due to stochastic forcing, $a_{\rm SF}$, is calculated using the following equations from \citet{Kasdin1995,Rein2010PhD}:
\begin{equation}
    a_{\rm SF} = a_{\rm SF} * \exp \left(\frac{-dt}{\tau}\right)
\end{equation}
\begin{equation}
    a_{\rm SF} = a_{\rm SF} + \left(\beta a_{\rm cent}\ X_i \ \sqrt{1 - \exp\left(\frac{-2\ dt}{\tau}\right)}  \right)
\end{equation}
where, $dt$ is the change in time from the previous step, $\tau$ is the auto-correlation timescale, $X_i$ is a randomly generated scaling factor from a normal distribution with mean equal to 0 and standard deviation equal to 1,  and the amplitude is determined by $\beta a_{\rm cent}$. In this prescription,  $a_{\rm cent} = Gm_{\rm AB}/\left(3a_{\rm bin}\right)^2$ is the gravitational acceleration due to the central stars at roughly the stability limit ($\approx 3a_{\rm bin}$),   and $\beta$ is a parameter that scales the amplitude of the acceleration due to turbulence relative to the acceleration due to the binary\footnote{\citet{rein2012PERIODRATIOS} used $\alpha$ to label this stochastic forcing scaled amplitude, but that notation may be confused with the classic \citet{shakura1973} $\alpha$ turbulence parameter, and hence we rename it as $\beta$.}.  The timescale  of the turbulence is typically the planet's orbital period. We set it to $5P_{\rm bin}$ since turbulence will be maximised near the disc truncation, at roughly this period.

 The stochastic forcing acceleration $a_{\rm SF}$ is independently calculated and added to both the $a_x$ and $a_y$ acceleration components of the planet. This means that stochastic forcing can change both $a_{\rm p}$ and $e_{\rm p}$. \citet{rein2012FOURTHREE,rein2012PERIODRATIOS} differed slightly by applying to the radial and azimuthal directions rather than $x$ and $y$. Our choice was driven by being computationally faster, for what should produce a qualitatively similar result.

The amplitude of stochastic forcing is only roughly constrained in the literature. \citet{Rein2009} estimated that $\beta\sim5\times10^{-6}$ for low-mass planets embedded in discs affected by MRI turbulence. \citet{rein2012PERIODRATIOS} tested $\beta = 10^{-5} - 10^{-7}$ in the context of reproducing the period ratio distribution in multi-planet single star systems, where there are some resonant pairs but not the ubiquity expected from smooth, convergent migration. It was found that $\beta = 10^{-5}$ disrupted all resonant chains, and hence was likely too high. We too test different values of $\beta$. 

We predict two potential effects of stochastic forcing. First, stochastic changes to $a_{\rm p}$ may have a stabilising effect by knocking a planet out of a mean motion resonance with the binary, which otherwise may have ejected the planet. This has been used by \citet{Adams2008,Ketchum2011,rein2012PERIODRATIOS} to explain the rarity of resonant chains in exoplanet systems. On the other hand, stochastic forcing will likely add eccentricity to an otherwise circular orbit. It was determined in Sect.~\ref{subsec:fully_unstable_resonances} that none of the known planets had to pass through a fully unstable resonance. However, a caveat to that argument was that some higher order resonances, whilst not fully unstable, would be unstable if the planet's eccentricity were high enough. Stochastic forcing may therefore imperil the planet if it increases its eccentricity when near resonance.

\section{Demonstrating The Effects}\label{sec:demonstration}

\begin{figure*}

    \includegraphics[width=0.8\textwidth]{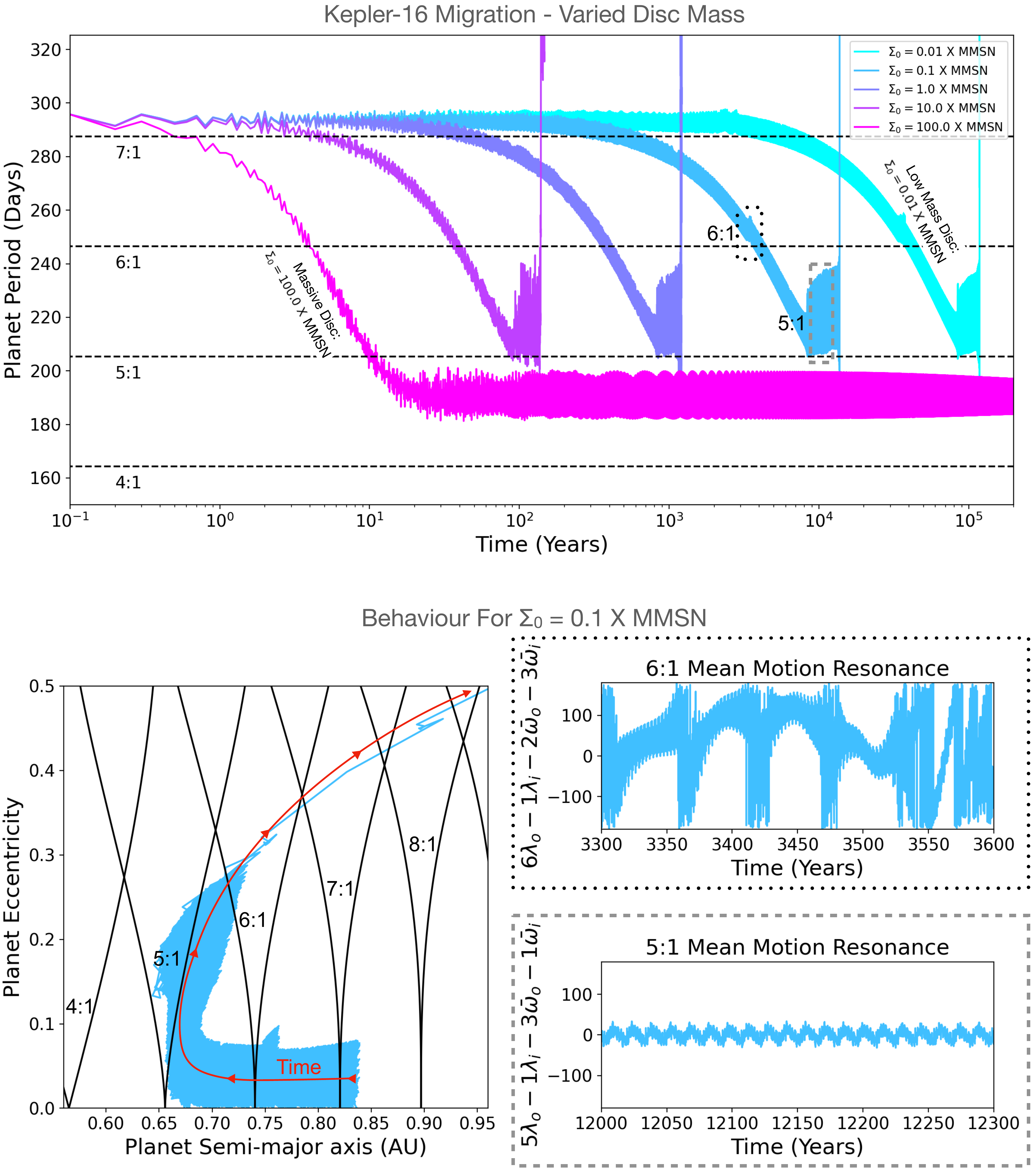}
    \caption{{\bf Top panel:} N-body migration simulations of a Kepler-16-like system but the parking location is at $0.62$ AU, which is closer than the observed $0.70$ AU location and interior to the $5:1$ resonance. In each simulation the only change is the disc mass, manifested by $\Sigma_0$, which we set between 100 and $0.01\times1700$ g/cm$^2$, where $1700$ g/cm$^2$ corresponds to a Minimum Mass Solar Nebula (MMSN). Stochastic forcing (simulating disc turbulence) is turned off. The parking time is essentially a linear function of $\Sigma_0$. Only for the most massive disc ($100\times$ MMSN) does the planet migrate quickly enough to pass through the $5:1$ resonance. All other slower planets are ejected at this resonance. No planets are ejected at the $6:1$ resonance, but the two slowest cases show a small amount of eccentricity excitement when passing there. {\bf Bottom panels:} Highlighted behaviour of the $0.1\times$ MMSN simulation, with the $a_{\rm p}$, $e_{\rm p}$ trajectory through the resonances from Eq.~\ref{eq:resonance_width} {\bf (left)} and the $5:1$ and $6:1$ resonant arguments from Eq.~\ref{eq:resonant_argument} {\bf (right)}. In the $6:1$ case all six resonant angles (including the one shown) circulate, with a period close to Eq.~\ref{eq:libration_period}. In the $5:1$ case this (and only this) angle librates, until the planet is ejected.}
    \label{fig:demo_running_gauntlet}
\end{figure*} 

\begin{figure*}

    \includegraphics[width=0.99\textwidth]{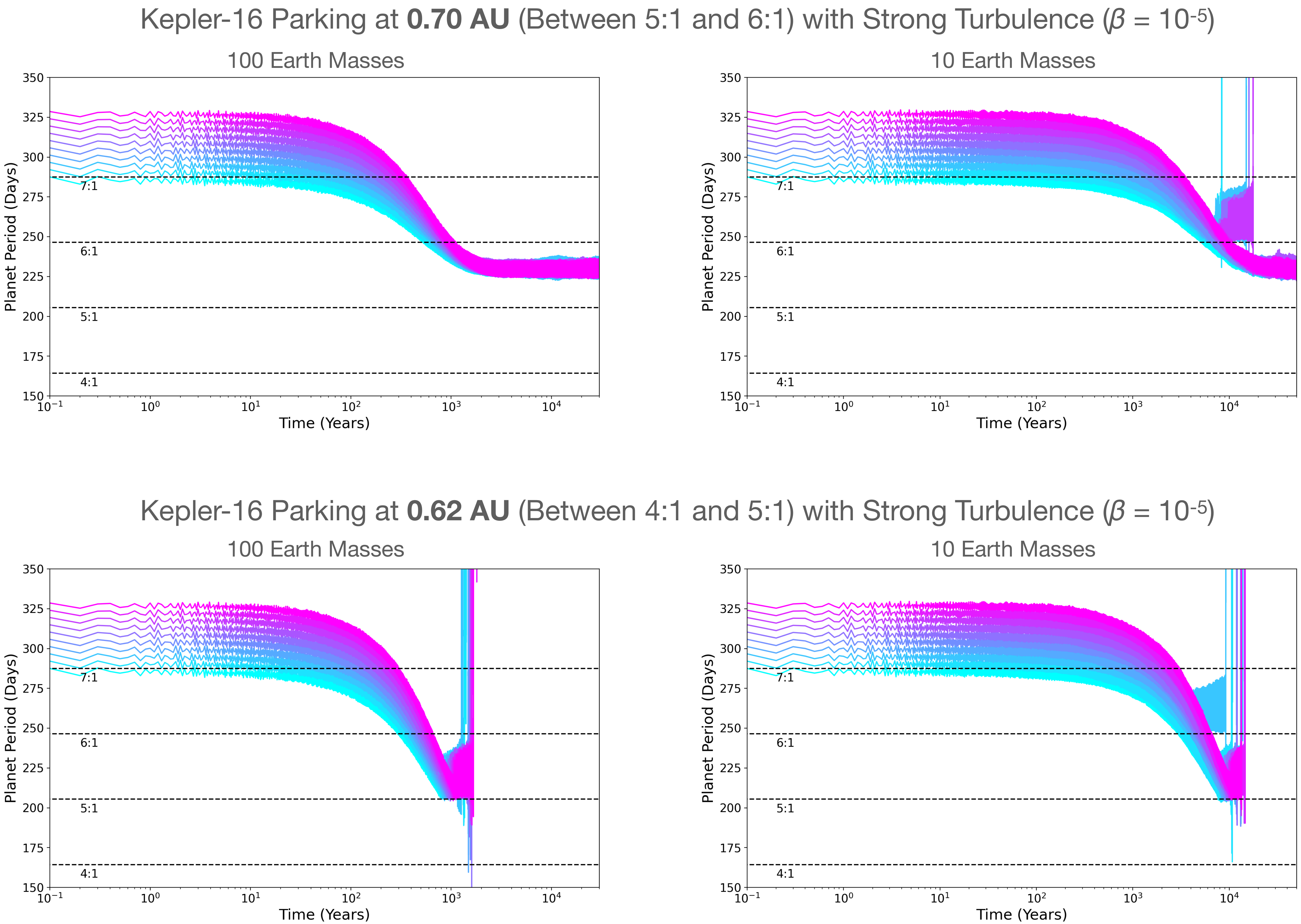}
    \caption{Demonstration of stochastic forcing, which is a simple method to model the effect of disc turbulence on planet migration. In all simulations we use the 41-day Kepler-16 binary, disc properties corresponding to $1\times$ MMSN and stochastic forcing is quantified by $\beta=10^{-5}$, which is considered strong turbulence by \citet{rein2012PERIODRATIOS}. In a given panel we test ten planets which are identical except for a slightly different starting location between 7 and 8 times the binary period. {\bf Top:} parking location at $0.7$ AU, corresponding to the known planet location. {\bf Bottom:} parking location at $0.62$ AU, located interior to the fully unstable $5:1$ resonance, requiring the planet to ``run the gauntlet''. {\bf Left:} $M_{\rm p}=100M_{\oplus}$ (like the known planet). {\bf Right:} $M_{\rm p}=10M_{\oplus}$. The top panels show that stochastic forcing is more likely to eject less massive planets. The bottom panels show that stochastic forcing is unlikely to help a planet pass through an unstable resonance.}
    \label{fig:stochastic_forcing_demonstration}
\end{figure*}

\begin{figure*}

    \includegraphics[width=0.99\textwidth]{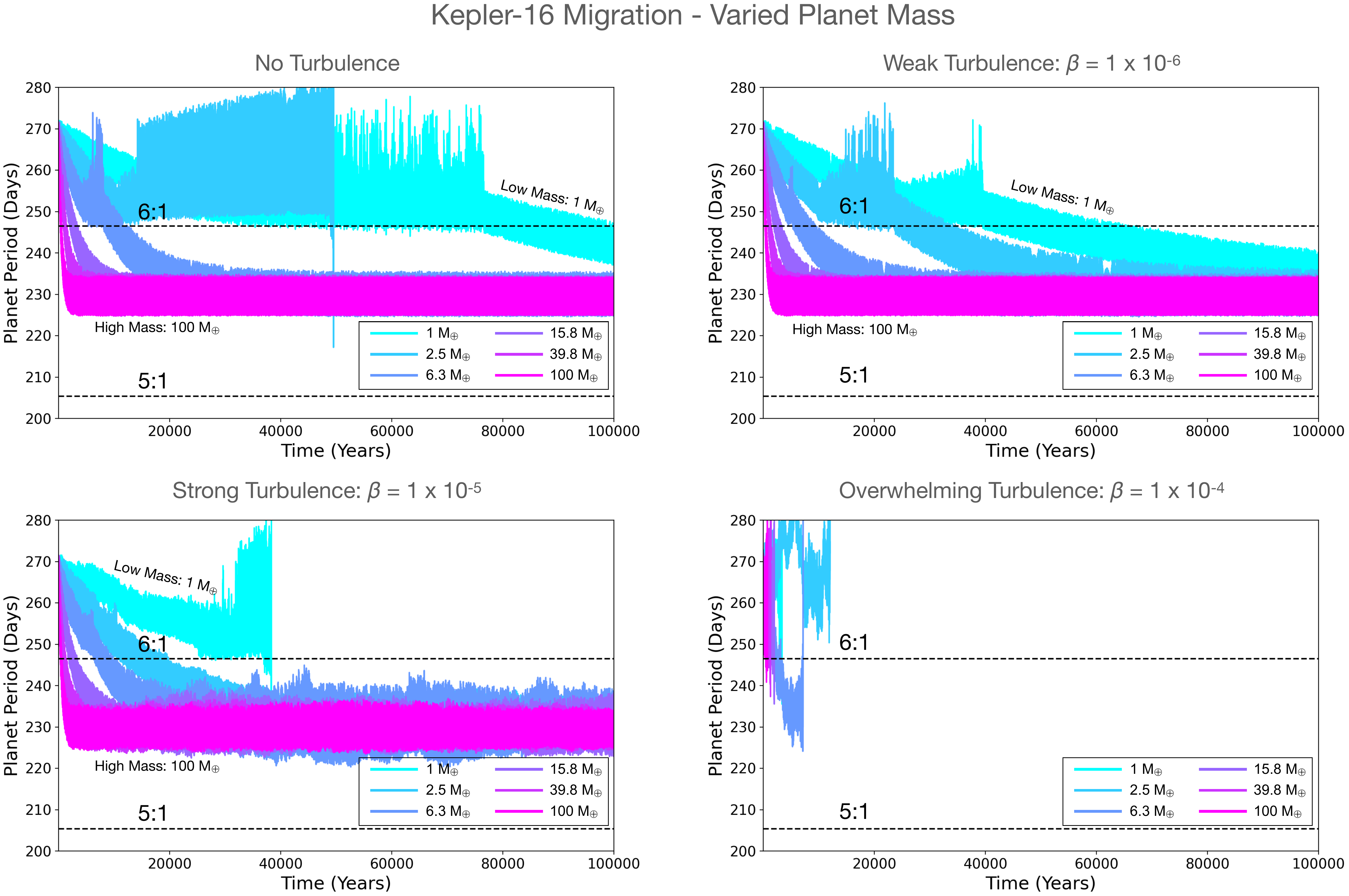}
    \caption{ Demonstration of Kepler-16 parking with different planet masses. Each panel has a different value of disc turbulence (modelled using stochastic forcing), including an ``overwhelming'' value where all planets go unstable, irrespective of mass. Note that unlike some of the previous figures, a linear time scale is used to better show how long each planet gets trapped in the $6:1$ resonance.}
    \label{fig:varied_planet_mass}
\end{figure*}

\subsection{Migration and parking}\label{subsec:simple_demo}

We first demonstrate our simple model of circumbinary planet migration and parking near the disc inner edge. We use Kepler-16 as an example, where a $100 M_{\oplus}$ planet starts at 1.37 AU (more than 15 times the binary period, well beyond any important resonances) and migrates in over 10,000 years, parking at the observed semi-major axis at 0.7 AU. The results are shown in the bottom panel of Fig.~\ref{fig:demo_simple}.

For the disc model we use $\Sigma_0=1700$ g/cm$^2$, corresponding to a Minimum Mass Solar Nebula (MMSN), scale height $h=0.04$, and $R_{\rm gap} = 2.86\times a_{\rm bin} = 0.64$ AU. Note that the factor of 2.86 is calculated such that the Lindblad and co-orbital torques balance at $a_{\rm p}=0.7$ AU. Stochastic forcing is turned off for this example.

The top plot of Fig.~\ref{fig:demo_simple} shows the profile of the disc surface density ($\Sigma$, Eq.~\ref{eq:disc_profile}) and the two torques: Lindblad ($T_{\rm L}$, Eq.~\ref{eq:lindblad}) and co-orbital ($T_{\rm co}$, Eq.~\ref{eq:coorbital}), where for clarity all profiles have been scaled between 0 and 1. The point of torque balance is seen to be slightly interior to the disc density peak, as was seen in hydrodynamical simulations (e.g. Fig. 5 of \citealt{thun2018}).

The middle plot of Fig.~\ref{fig:demo_simple} shows the two torques acting on the planet over time. Initially, when the planet is farther out than $\sim1$ AU, the disc density profile is like that of an untruncated disc, where $\Sigma\propto a_{\rm p}^{-3/2}$. Since the Oort constant $B=\Omega_{\rm p}/4$ is also $\propto a^{-3/2}$ (Eq.~\ref{eq:orbital_frequency}), the two differentials in Eq.~\ref{eq:coorbital} cancel, leaving $T_{\rm co}\sim0$. The Lindblad torque on the other hand increases with $\Sigma$. As the planet moves closer than 1 AU the co-orbital torque has a steep increase due to the disc truncation, to the point of balancing the Lindblad torque at $0.7$ AU.  The parking location of the planet will always be slightly {\it interior} to the peak of the disc density profile.

The bottom plot of Fig.~\ref{fig:demo_simple} shows the product of our migration and parking model, with the Kepler-16 planet reaching its known semi-major axis after about 4000 years. Interestingly, if we compare this to the seminal paper \citet{pierens2013} we see that both the migration behaviour and timescale are similar to their hydrodynamical simulations (their Fig. 4).


\subsection{Running the gauntlet}\label{subsec:gauntlet_demo}

In this section we take the same simulation of a $100M_{\oplus}$ planet around Kepler-16 from Fig.~\ref{fig:demo_simple} but make three modifications. First, we reduce $R_{\rm gap}$ from $2.86\times a_{\rm bin}$ to  $2.52\times a_{\rm bin}$, which will attempt to park the planet at $0.62$ AU, which is interior to the fully unstable $5:1$ resonance and on a small island of stability (Fig.~\ref{fig:stability_map}). Whilst in Sect.~\ref{subsec:fully_unstable_resonances} it was determined that no planets (including Kepler-16) existed interior to fully unstable resonances, we still want to test if a planet could hypothetically pass such a resonance if migrating quickly enough. 
Second, we vary $\Sigma_0$ over five orders of magnitude, between $100 \times $ MMSN (i.e. $1700\times 100$ g/cm$^2$) and $0.01\times $ MMSN. The ``parking time'' of a planet will vary approximately linearly with the total disc mass (characterised in our equations by $\Sigma_0$). We will therefore see if a  massive disc can precipitate sufficiently fast migration to cross the $5:1$ resonance. On the other hand, this also tests if slow migration in a low-mass disc can imperil even the $6$ or $7:1$ resonance.

A third but small change is that the planet starts closer, just outside the $7:1$ resonance. This is done for computational expediency, particularly for the slowly-migrating systems. As long as the planet starts beyond any potentially destabilising resonances, the starting point should be irrelevant.

The results are shown in Fig.~\ref{fig:demo_running_gauntlet}. The top panel shows the planet's period over time, where the log scale demonstrates how a change in the order of magnitude of $\Sigma_0$ changes the order of magnitude of the parking time. For 100 $\times$ MMSN the planet can indeed run the gauntlet and pass the unstable $5:1$ resonance and park on a stable orbit. For all less massive discs the planet migrates too slowly and is captured into a $5:1$ resonance and ejected on a similar timescale in each case (on a log scale). 

On the other hand, the higher order $6:1$ and $7:1$ resonances remain stable regardless of the disc mass, and hence migration rate. This matches the conclusion of Sect.~\ref{subsec:fully_unstable_resonances} that these resonances are stable as long as the planet's eccentricity remains sufficiently low.

The bottom panels highlight the behaviour for a $0.1\times$  MMSN disc. On the left we show the evolution of $a_{\rm p}$ and $e_{\rm p}$, where time passes roughly following the arrow. Resonant width wedges from Eq.~\ref{eq:resonance_width} taken from \citep{mardling2013} are overlayed. Throughout the migration there is always eccentricity variation of magnitude a bit less than $0.1$, due to eccentricity pumping from the binary \citep{leung2013}. There is interestingly a ``flutter'' of additional variation around the $6:1$ resonance, albeit offset from the exact commensurability. Nevertheless, the planet passes through unscathed. The $5:1$ resonance, however, was a bridge too far and we see the simultaneous increase of both $e_{\rm p}$ and $a_{\rm p}$ to ejection.

The bottom right panels highlight 300 year snapshots of the planet's behaviour at $5$ and $6:1$ resonances, by showing one of the resonant arguements for each. For the $6:1$ resonance this argument (and the five others) are not librating, so the planet does not officially enter the resonance. Its circulation has a period matching the 58 years which one can calculate from Eq.~\ref{eq:libration_period}, using $e_{\rm p}=0.05$. For the $5:1$ resonance this argument is indeed librating, with a period matching the 15 years from Eq.~\ref{eq:libration_period}. This short libration period leads to the planet's resonant capture and ultimate ejection.

Overall, in this simple demonstration we see two things. One, a fast migration rate may allow planets to pass through fully unstable resonances. Two, a slow migration does not necessarily imperil planets at otherwise stable resonances.

\subsection{Stochastic forcing}

Neither simulation in Sect.~\ref{subsec:simple_demo} or ~\ref{subsec:gauntlet_demo} contained stochastic forcing. As a demonstration we take the Kepler-16 system with planets started between the $7:1$ and $8:1$ resonance, use a 1 $\times$ MMSN disc and add stochastic forcing with a high level of turbulence ($\beta=10^{-5}$). We test planets with $100$ and $10M_{\oplus}$, where the former roughly matches the known planet mass. We set $R_{\rm gap}$ to 2.86 and 2.52 AU, which will make the planet park exterior (0.7 AU) and interior (0.62 AU), respectively, to the fully unstable $5:1$ resonance. The results are shown in Fig.~\ref{fig:stochastic_forcing_demonstration}.

The only effect of the starting location should be to provide a ``random seed'' for the stochastic forcing. When parking at 0.7 AU, we see that the $100M_\oplus$ planet is stable in all ten simulations, but for the $10M_\oplus$ planet some simulations have an ejection at the $6:1$ resonance. This instability is a result of eccentricity pumping from the turbulence. Whilst the amplitude of stochastic forcing (and hence eccentricity variations) is independent of planet mass, the time spent near this resonance is dependent of planet mass, which makes ejection of the smaller planet more likely. We also see that this effect is indeed stochastic, with only four of the ten $10M_{\oplus}$ planets being ejected.

When parking at 0.62 AU, we see that all of the $100M_{\oplus}$ planets are ejected at the $5:1$ resonance; they are unable to run the gauntlet. This instability is the same result as Fig.~\ref{fig:demo_running_gauntlet} for $\Sigma_0$ without stochastic forcing. We see that stochastic forcing, even at this high level, was unable to save the migrating planet. Unsurprisingly, all $10M_{\oplus}$ planets are ejected too, stochastically at either the $5:1$ or $6:1$ resonance.

 In Fig.~\ref{fig:varied_planet_mass} we show an additional demonstration of how a planet's mass affects its evolution in a turbulent disc. Using Kepler-16 as an example, we migrate a planet through the potentially treacherous $6:1$ resonance until parking at $\approx228$ days, testing four different values of turbulence and six different planet masses (log-uniform between 1 and $100M_\oplus$). We see that for no turbulence (top left) the three least massive planets get trapped in a $6:1$ resonance, although only one is ejected. When weak turbulence ($\beta=10^{-6}$) is added we interestingly see that now all planets are stable. It is possible that in select cases turbulence could actually help break potentially destabilising resonances, although we would still conclude overall that turbulence is typically detrimental to a planet's survival. For strong turbulence ($\beta=10^{-5}$) only the smallest planet is ejected, but we do see a greater amount of orbital variation of parked planets (which is greater at low mass). Such variation so close to the stability limit could potentially lead to future instability. 

Finally, we also test $\beta=10^{-4}$, which is an order of magnitude beyond the bounds tested by \citet{rein2012PERIODRATIOS}. We see that all planets are quickly ejected, irrespective of planet mass. Such turbulence is unrealistic since we do know that at least giant planets can survive around binaries, so this gives us a rough upper limit to $\beta$, in line with previous studies.

\subsection{Varied eccentricity damping}\label{subsec:eccentricity_damping}

\begin{figure}

    \includegraphics[width=0.49\textwidth]{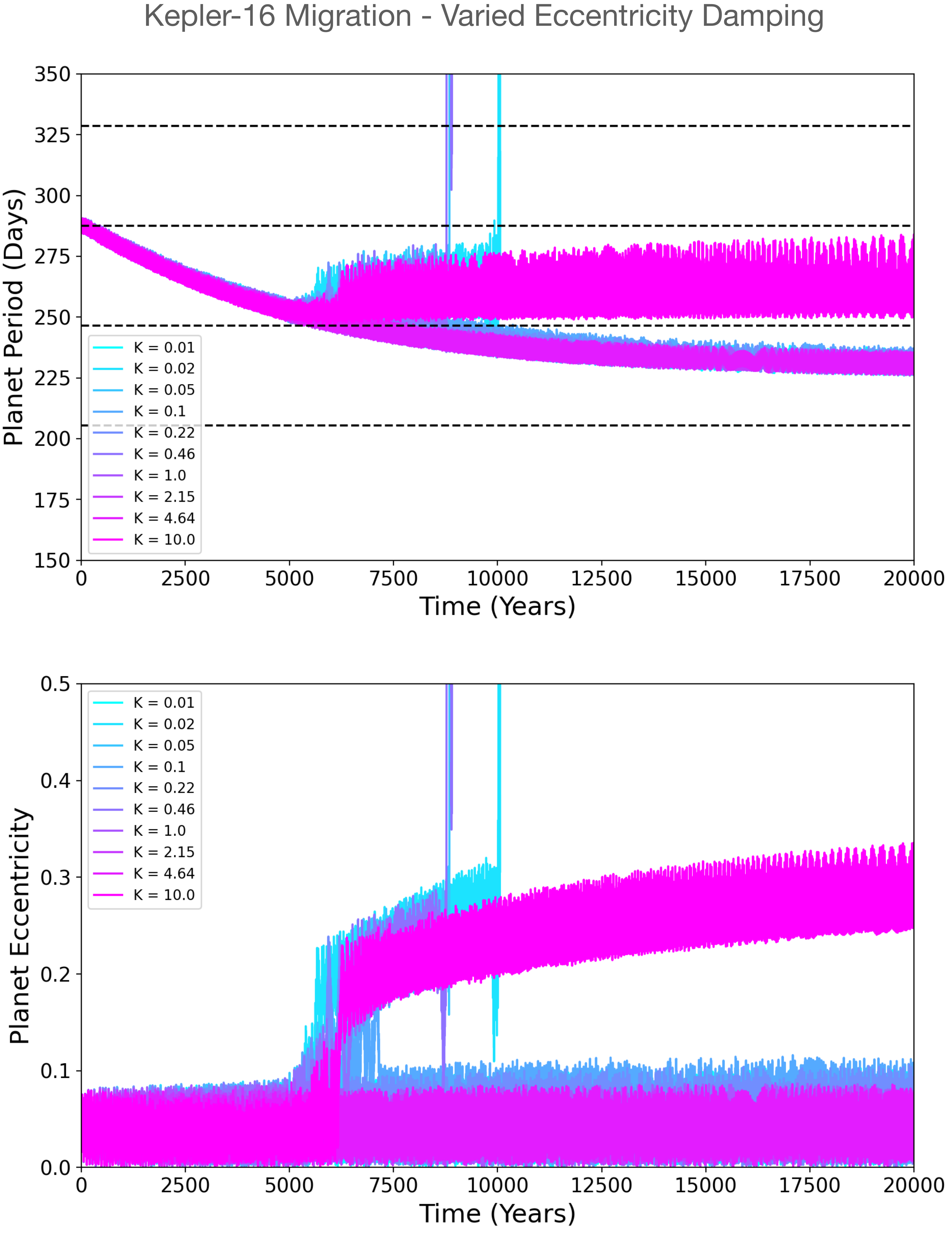}
    \caption{ Migration and parking of a $10M_{\oplus}$ Kepler-16b with different levels of eccentricity dampening. In this paper we use the typical value of $K=10$ \citep{lee2002,Kley2004}, but the bluer colours test lower eccentricity dampening. There is no obvious trend of planet ejection being more common with lower eccentricity damping, at least for this example passing through the $6:1$ resonance which is partially unstable.}
    \label{fig:K_variation_paper_plot}
\end{figure}

 The rate of eccentricity damping is related to the rate of semi-major axis damping (i.e. the migration rate) by $K$. In Fig.~\ref{fig:K_variation_paper_plot} we test lesser eccentricity damping down to $K=0.01$. Stochastic forcing is turned off for this example. We use a lower mass Kepler-16 example at $10M_\oplus$ since a smaller planet will migrate slower and may be at higher risk. We see in four examples the planet gets caught in a $6:1$ resonance, three of which are ejected within 20,000 years. This occurrence though is seemingly random, with no clear trend with a lower $K$. In the eccentricity plot we can see that there is higher variation for the blue curves, i.e. with a lower $K$, but this seemingly does not make ejection more likely, at least in this example.

\begin{figure*}

    \includegraphics[width=0.99\textwidth]{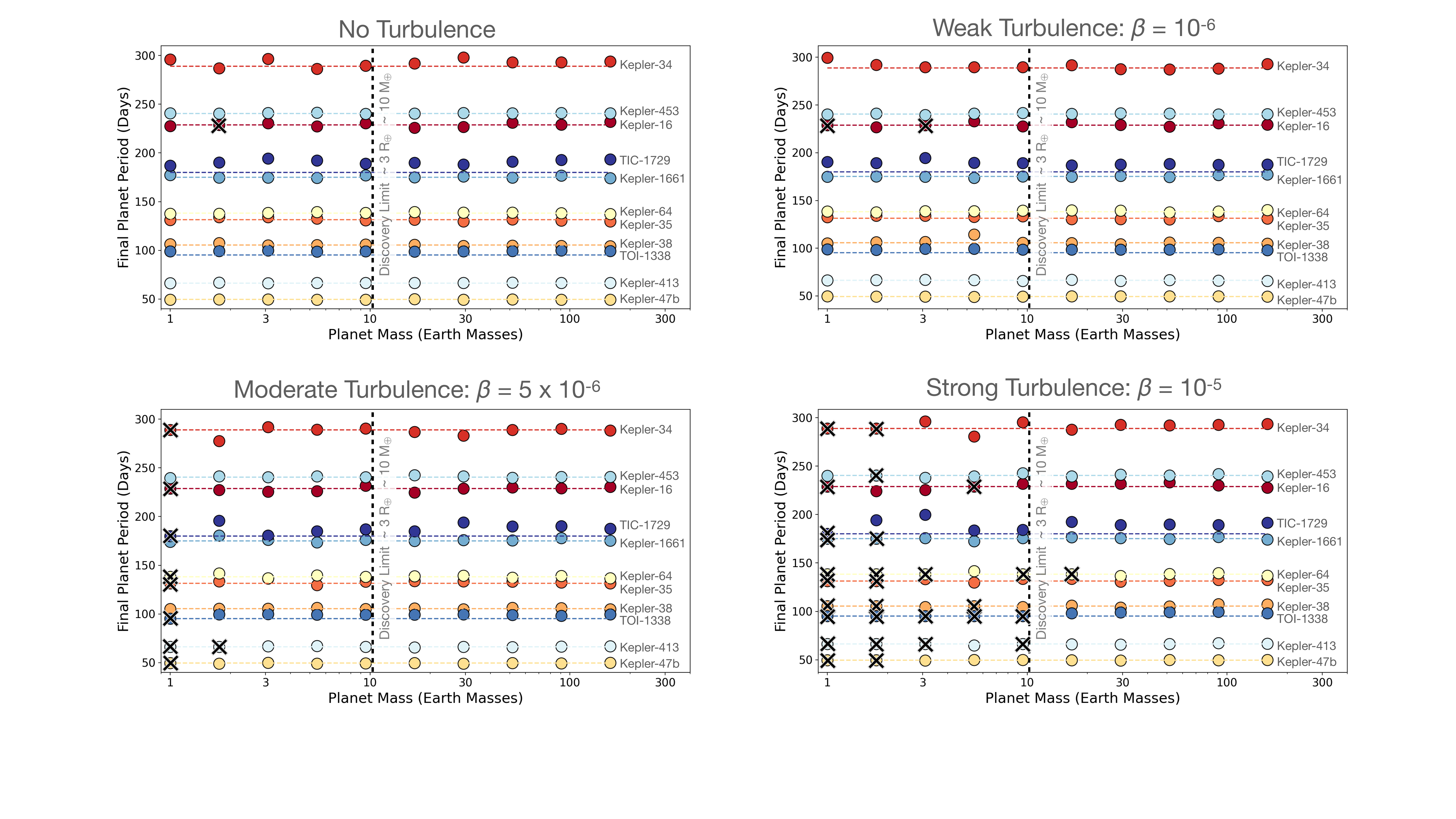}
    \caption{Survival of migrating planets of different masses around the known circumbinary planet hosts. The dashed lines indicate the period of the known planet, and this is the orbit to which the planet is migrating. A circle indicates the final period in the simulation if the planet remained stable. A crossed out circle indicates that that particular planet mass went unstable during its migration. Colours are solely for visual clarity to distinguish between binaries. Vertical lines in each panel denote the rough discovery limit at $3R_{\oplus}\sim 10M_{\oplus}$, which is based on a very rough mass-radius relation \citep{Seager2007,Bashi2017}. The four panels have different levels of disc turbulence (stochastic forcing,  quantified by $\beta$, the amplitude of stochastic forcing acceleration relative to the acceleration from the central binary). These results show that resonant ejection of migrating planets does exist and targets small planets, but a high amount of disc turbulence is required for the effect to be applicable beyond the smallest (Earth-mass) planets.}
    \label{fig:big_sim}
\end{figure*}

\section{Application To Known Planets}\label{sec:applications}

We conduct a test to see if the known circumbinary planets could have safely migrated to their observed locations if they had a lower mass. For each binary\footnote{With the exception of Kepler-1647.} we calculate the known planet - binary period ratio, round it up to the nearest integer, add 1.3 to this value and set this as the starting planet - binary period ratio. Mathematically,  $P_{\rm p} = [1.3 + {\rm Ceil} (P_{\rm p}/P_{\rm bin})] \times P_{\rm bin}$.  This means that for consistency, every tested planet has to migrate through two mean motion resonances to reach a parking location matching the known planet. For example, Kepler-16 has $P_{\rm p} / P_{\rm bin} = 5.57$ so it starts at $7.3\times P_{\rm bin} = 299$ days. The integration time is scaled with planet mass, $t_{\rm int} = 1,000,000  ({\rm yr}) / M_{\rm p}  ({\rm M_{\oplus})}$, allowing all planets to reach their parking location irrespective of migration speed.

For all binaries we test ten planet masses uniformly log-spaced between $1M_{\oplus}$ and $159M_{\oplus} = 1/2M_{\rm Jup}$, a rough range where Type-I migration should be applicable. For the disc parameters we use a MMSN ($\Sigma_0=1700{\rm g/cm}^2$), scale height $h=0.04$ and $R_{\rm gap}$ is calculated in each simulation such that the Lindblad and co-orbital torque balance occurs at the observed planet period. The main determinant of the planet's survival will be eccentricity pumping from stochastic forcing, which mimics disc turbulence. We run one test without stochastic forcing, and then three with different values of $\beta = 10^{-6}$, $5\times10^{-6}$ (derived value by \citealt{Rein2009}) and $10^{-5}$ (considered  strong, even to the point of being potentially unrealistically high by \citealt{rein2012PERIODRATIOS}).

In Fig.~\ref{fig:big_sim} we show the results of these simulations by plotting the final planet period as a function of planet mass for all 11 tested binaries and all four different levels of disc turbulence. For simulations where the planet was ejected we plot a crossed-out circle centered on the period of the known planet (dashed line). A vertical line is placed at $10M_{\oplus}$ in all of the panels, which roughly corresponds to our present transit discovery limit of $3R_{\oplus}$, although we note that there exists significant scatter in the exoplanet mass-radius relationship \citep{Seager2007,Bashi2017}.

In the case of no turbulence, we see that all but one simulation manages to park near the known period and retain a stable orbit. The final planet periods show some scatter around the known period (the dashed lines). This is an expected consequence of the three-body interactions with the binary causing the instantaneous osculating period of the planet to vary by about $5\%$ \citep{leung2013,armstrong2013}. The one unstable simulation corresponds to a $1.75M_\oplus$ Kepler-16 planet, which like in Fig.~\ref{fig:varied_planet_mass} was not the lowest mass planet, but this is likely a coincidence. In this case eccentricity-pumping from the binary alone was sufficient to induce instability at the $6:1$ resonance.  In Fig.~\ref{fig:stability_map} we see that for Kepler-16b to pass the $6:1$ resonance without any chance of ejection (i.e. to stay in the dark blue stable region) its eccentricity needs to stay below $\approx0.03$. This is a tight limit, and hence it is not surprising for an occasional case where a planet is ejected even in the absence of turbulence. Aside for this outlier, we see agreement with the Sect.~\ref{subsec:fully_unstable_resonances} conclusion that none of the known planets had to pass through fully unstable resonances.

When stochastic forcing is turned on at a weak level ($\beta=10^{-6}$) there is only a small change relative to the no turbulence simulations, with now two low-mass Kepler-16 planets being ejected instead of one. At a moderate level of turbulence 7/11 of the $1M_{\oplus}$ planets are ejected, but only one more massive planet is ejected. Only at the strongest tested level of turbulence do we see a significant ejection of small planets, with 10/11 $1M_{\oplus}$ planets becoming unstable. When broadening the range to higher masses, $67\%$ of $\leq 3M_{\oplus}$ are ejected and $51\%$ of $\leq 10M_{\oplus}$ are ejected. 

\begin{figure}

    \includegraphics[width=0.49\textwidth]{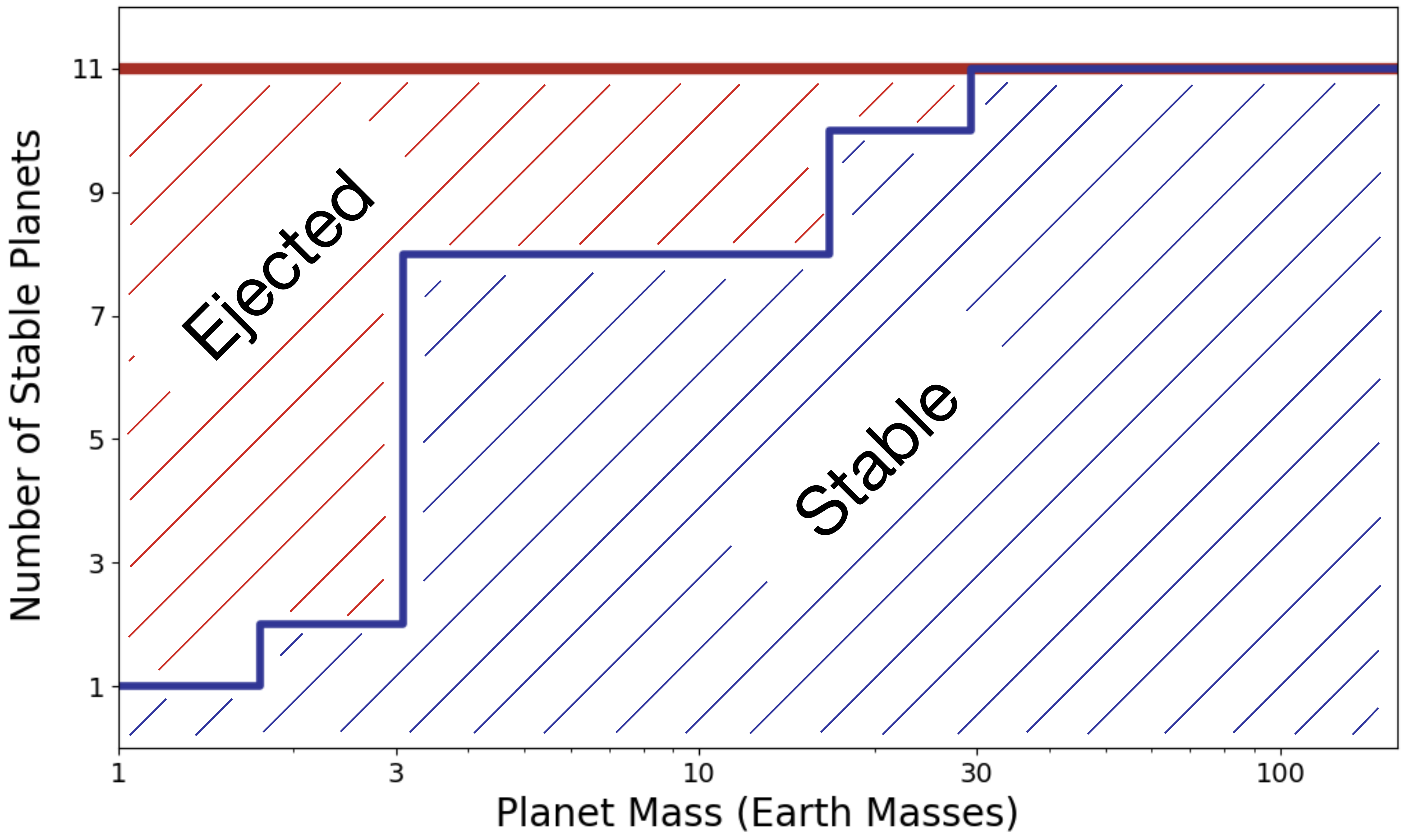}
    \caption{Sculpting of the circumbinary planet size distribution by resonant ejection. The blue curve is a count of the 11  Kepler/TESS planets that tested in Fig.~\ref{fig:big_sim} that survived migration through strong turbulence ($\beta=10^{-5}$), as a function of planet mass. The red "ejected" region corresponds to the black crosses in Fig.~\ref{fig:big_sim}. Resonant ejection preferentially targets small planets, but does not remove all of them.}
    \label{fig:size_distribution}
\end{figure}

 In Fig.~\ref{fig:size_distribution} we show how our starting log-uniform distribution of planet masses between $1M_\oplus$ and $159M_\oplus$ is sculpted over the course of the N-body simulations. We only show results in the case of strong turbulence ($\beta=10^{-5}$), i.e. this is the maximum efficiency for this effect. Even in this extreme case of a highly turbulent disc, resonant ejection  is not efficient enough to account for a complete dearth of small circumbinary planets.

\section{Discussion}\label{sec:discussion}

\subsection{Circumbinary planet size distribution}




At present, the observed size distribution of planets around one and two stars is substantially different; around single stars a plethora of super-Earths and mini-Neptunes have been discovered, whereas around binaries the smallest planet to date is Kepler-47b at $3R_{\oplus}$ (Fig.~\ref{fig:known_planets}, right). \citet{armstrong2014} made the first circumbinary planet abundance calculations and showed that within the gas giant regime it was roughly similar to that around single stars, but no inferences could be made below $4R_{\oplus}$.  The observational bias is that the transit timing variations for circumbinary planets are on the order of days \citep{armstrong2013}, and hence larger than the typical transit durations (which also vary, for added complexity, \citealt{kostov2014}) . This means that the known sample came from by-eye detections, which required individually deep transits and hence were restricted to large planets. Since then  new techniques for finding small transiting circumbinary planets  have been developed, which can phase-fold photometric data on a variable timing and duration to reveal shallow transits of small planets  \citep{windemuth2019b,martin2021}. They are yet to be widely applied though. We therefore cannot yet say that the absence of small circumbinary planets truly reflects reality, and we can only speculate on the existence of circumbinary super-Earths.

 In this paper we propose a theoretical explanation of a dearth of small circumbinary planets: resonant ejection precipitated by slow migration. However, our numerical simulations in Sect.~\ref{sec:applications} demonstrate that this effect is only efficient if there is a significant means of raising the planet's eccentricity. We propose that disc turbulence could cause this, but even at what is considered a large degree of turbulence only roughly half of the small ($\lesssim 3R_\oplus$) planets are ejected.

Aside for observational limitations, {\it what other theoretical arguments could be made for an absence of small circumbinary planets?} \citet{fleming2018} and \citet{sutherland2019} proposed that the tidal evolution of tight ($\lesssim 7$ days) binaries may induce late-stage ejection of planets that have finished migrating. They did not test a dependence on planet mass though.  Another possibility is that the presence of a second circumbinary planet could cause scattering that ultimately results in a planet being ejected. In a single-star, two-planet system, \citet{chatterjee2008} showed that it is more likely that the smaller planet would be ejected . \citet{sutherland2019} demonstrated that in a binary system the complex interplay of planet-planet and binary-planet mean motion resonances may increase instability. For example, a planet's eccentricity may be raised by a companion planet, bringing it into the resonance overlap region with the binary. In the upcoming Fitzmaurice, Martin \& Fabrycky (in prep.) we investigate how the size distribution of circumbinary planets is sculpted in multi-planet systems, providing a complementary effect to the single-planet resonant ejection tested in this paper. 

Finally, in the context of single stars, there is an on-going debate about whether the large populations of transiting super-Earths ($P_{\rm p}\lesssim 100$ days) formed in situ \citep{Hansen2013,Chiang2013,Lee2016,MacDonald2020} or via migration \citep{Terquem2007,Ogihara2015,Inamdar2015,Raymond2018,Izidoro2021}. For circumbinary planets in situ formation so close to a binary is likely prohibited due to disc truncation \citep{Artymowicz1994,miranda2015,ragusa2020} or high turbulence \citep{Paardekooper2012,Lines2014,Meschiari2014,pierens2020,Pierens2021}. It is therefore possible that super-Earths do not exist around binaries simply because their preferred in situ formation path is quenched. We encourage future studies of comparative single-binary population synthesis to test this possibility.

\subsection{Assumptions and caveats}\label{subsec:discussion_caveats}

We propose and test a mechanism to explain a dearth of small circumbinary planets by a slow migration through mean motion resonance. There are two aspects to our test. First, we demonstrate using N-body stability maps in Sect.~\ref{subsec:fully_unstable_resonances} that none of the known planets exist interior to fully unstable resonances. The conclusion from this part of our work is that all of the known planets could have migrated to their known location independent of their speed, and hence independent of their mass, as long as they maintained a sufficiently low eccentricity. This aspect of our research, which would argue against  the prevalence of resonant ejection of slowly-migrating planets, should be relatively uncontroversial, as it comes from simply three-body stability tested using N-body integrations. A potential caveat may be that we assumed purely coplanar systems, although all of the known planets are aligned to within a few degrees of the binary \citep{martin2017} and there should not be a strong stability/inclination dependence in such cases \citep{Pilat-Lohinge2003,Doolin2011,martin2016}. We also only assume single planets, whereas multi-planet systems in binaries introduce complex dynamics, as seen by \citep{Hinse2015,quarles2018,sutherland2019} and investigated in our upcoming work Fitzmaurice, Martin \& Fabrycky (in prep).

The second aspect of our work was to construct N-body simulations of circumbinary planet evolution, using simple prescriptions for truncated discs, migration, parking and stochastic turbulence. Some of these pieces of our research have been tested in previous studies but only for single stars (e.g. the torque prescriptions from \citealt{tanaka2002,lubow2010} and stochastic forcing from \citealt{rein2012FOURTHREE}). N-body circumbinary migration was used in \citealt{sutherland2019} but only  constant inwards migration without a parking mechanism. Our truncated disc profile and physics of planet parking near the inner edge are based on and qualitatively match the hydrodynamical simulations of \citet{Pierens2008,pierens2013,thun2018,Penzlin2021}, but our simpler simulations will inevitably lack some subtitles. 

In our work we only consider Type-I migration, which is only applicable to small planets embedded in a disc. Larger planets are expected to carve a hole in the disc and migrate at a different (typically slower) Type-II rate. We justify this assumption not only for small planets but for all of the known planets based on \citet{Pierens2008}. Before any planets were discovered, they predicted that if the planet were in the Type-II regime then the massive planet would clear out its co-orbital region, suppressing the co-orbital torque needed to oppose the inwards Lindblad torque (Fig.~\ref{fig:demo_simple}) and prevent the planet from getting too close to the binary. This may be why no Jupiter-mass or higher planets have been found near the stability limit, and it suggests that all known planets near the stability limit must have remained in the Type-I regime.  Another explanation, as suggested by \citet{sutherland2019}, is that Type-II migration precipitated ejection due to its slow migration rate, similar to what we see for small planets given sufficient turbulence.

Even if the Type-I assumption is reasonable, our work is still underpinned by a fundamental assumption that all Type-I planets will park near the disc inner edge irrespective of their mass. This parking comes from a torque balance between the inwards Lindblad torque ($T_{\rm Lind}\propto \Sigma$) and the outwards co-orbital torque ($T_{\rm co}\propto d\Sigma/dr$). Using hydrodynamical simulations to test if this underlying physics holds down to Earth and super-Earth sized planets would be interesting. \citet{thun2018,Penzlin2021} did show that smaller planets park slightly farther from the binary, but the overall ability for a Type-I planet to park was not shown to have a mass dependence.  These studies also only covered  $M_{\rm p}>16M_\oplus$, i.e. larger than the observed dearth of small planets.

One potentially important effect which we do neglect is the role of disc eccentricity. Even a circular binary will induce eccentricity in the inner regions of the disc, and this eccentricity will increase for eccentric binaries \citep{Kley2015}. In our simulations the disc always acts to damp the planet eccentricity towards zero. This may be justified based on the low eccentricities of the known planets (all but two have $e_{\rm p}<0.1$).  On the other hand, \citet{thun2018,Penzlin2021} show that more massive planets tend to circularise the inner regions of the disc, whereas for smaller planets the disc remains eccentric. It may therefore be possible that an eccentric disc induces a small but non-negligble eccentricity in small planets, which may act to make resonances unstable even without the need for turbulence.

Overall, whilst we believe that our N-body work is sufficient to illuminate and provide a preliminary investigation of this effect, and was necessary for covering a broad parameter space within a computationally manageable time, we highly encourage  more hydrodynamical simulations of mass-dependent circumbinary migration to verify or challenge our results.

\section{Conclusion}\label{sec:conclusion}

There have been 14 circumbinary planets discovered. They are all large and gaseous ($>3R_\oplus$), and the majority of them are interspersed between potentially unstable first-degree mean motion resonances with the binary ($5:1$, $6:1$ and $7:1$). It is believed that the planets migrated through these resonances. Since planets migrate at a speed proportional to their mass, a possible explanation for a dearth of small circumbinary planets could be resonant ejection of a slow-migrating planet. In this paper we tested if this is a viable mechanism.

First, we used N-body simulations to show that that none of the 14 planets exist interior to a fully unstable resonance, which we define to be one for which even circular planets would be unstable. This means that migration to the locations of the known planets would be possible for planets of any migration rate, and hence mass, as long as their eccentricity remained low. 

Second, we constructed N-body simulations of migrating planets in a turbulent and truncated disc, using simplified but realistic prescriptions for Lindblad and co-orbital torques to mimic Type-I migration  and parking near the disc's inner edge. We demonstrated that with weak or no turbulence, indeed  almost any planet mass may survive,  with ejections being rare. As turbulence is increased, the planet's eccentricity may be increased to the level that high-order resonances become unstable. We also showed that this mechanism disproportionately effects small planets, due to their slow migration rate. 

Nevertheless, whilst the effect of resonant ejection may occur in nature, our extended simulations show that a potentially unrealistically high level of turbulence is required to make it applicable to more than just the smallest Earth-mass planets. It is therefore unlikely that this effect could explain a complete or significant dearth of $<3R_\oplus$ circumbinary planets. 

At present, this observed lack of small planets is likely an observational bias \citep{martin2018,windemuth2019b,martin2021}. If it is determined to be true though, it is likely that an alternate (or at least additional) explanation will be required for their absence. One possibility is the dynamics of multi-planet systems, within which the smallest planet is the most likely to be ejected. We investigate this in a separate work (Fitzmaurice, Martin \& Fabrycky, in prep). Alternatively, small planets (including super-Earths) may have a tendency to be formed in situ in the inner regions of protoplanetary discs, and this process is suppressed around binaries.

\section*{Data availability statement}

The observational data used in this paper is limited to the physical and orbital parameters of the known planets, which are all listed in Table~\ref{tab:known_planets}. The simulation data used in this paper is available on request by contacting the corresponding author.

\section*{Acknowledgements}

 We are greatly appreciative of our referee Anna Penzlin for a comprehensive review which undoubtedly improved the quality of this paper. We thank Dan Fabrycky for continued discussions on circumbinary planet dynamics over the years, and Rosemary Mardling for her expert comments on mean motion resonances. We also thank the Ohio State exoplanet research group for helpful comments when presenting earlier versions of this work, and we thank the Arizona star and planet formation research group for their insightful comments. Hanno Rein and Dan Tamayo were great helps in assisting our implementation of stochastic forcing in ReboundX. Support for this work was provided by NASA through the NASA Hubble Fellowship grant
\#HF2-51464 awarded by the Space Telescope Science Institute, which is operated by the Association of Universities for Research in Astronomy, Inc., for NASA, under contract NAS5-26555. This research has made use of the NASA Exoplanet Archive, which is operated by the California Institute of Technology, under contract with the National Aeronautics and Space Administration under the Exoplanet Exploration Program.

\bibliographystyle{mnras}
\bibliography{references}

\begin{thebibliography}{}
\makeatletter
\relax
\def\mn@urlcharsother{\let\do\@makeother \do\$\do\&\do\#\do\^\do\_\do\%\do\~}
\def\mn@doi{\begingroup\mn@urlcharsother \@ifnextchar [ {\mn@doi@}
  {\mn@doi@[]}}
\def\mn@doi@[#1]#2{\def\@tempa{#1}\ifx\@tempa\@empty \href
  {http://dx.doi.org/#2} {doi:#2}\else \href {http://dx.doi.org/#2} {#1}\fi
  \endgroup}
\def\mn@eprint#1#2{\mn@eprint@#1:#2::\@nil}
\def\mn@eprint@arXiv#1{\href {http://arxiv.org/abs/#1} {{\tt arXiv:#1}}}
\def\mn@eprint@dblp#1{\href {http://dblp.uni-trier.de/rec/bibtex/#1.xml}
  {dblp:#1}}
\def\mn@eprint@#1:#2:#3:#4\@nil{\def\@tempa {#1}\def\@tempb {#2}\def\@tempc
  {#3}\ifx \@tempc \@empty \let \@tempc \@tempb \let \@tempb \@tempa \fi \ifx
  \@tempb \@empty \def\@tempb {arXiv}\fi \@ifundefined
  {mn@eprint@\@tempb}{\@tempb:\@tempc}{\expandafter \expandafter \csname
  mn@eprint@\@tempb\endcsname \expandafter{\@tempc}}}

\bibitem[\protect\citeauthoryear{{Adams}, {Laughlin}  \& {Bloch}}{{Adams}
  et~al.}{2008}]{Adams2008}
{Adams} F.~C.,  {Laughlin} G.,   {Bloch} A.~M.,  2008, \mn@doi [\apj]
  {10.1086/589986}, \href
  {https://ui.adsabs.harvard.edu/abs/2008ApJ...683.1117A} {683, 1117}

\bibitem[\protect\citeauthoryear{Armstrong et~al.,}{Armstrong
  et~al.}{2013}]{armstrong2013}
Armstrong D.,  et~al., 2013, \mn@doi [Monthly Notices of the Royal Astronomical
  Society] {10.1093/mnras/stt1226}, 434, 3047

\bibitem[\protect\citeauthoryear{{Armstrong}, {Osborn}, {Brown}, {Faedi},
  {G{\'o}mez Maqueo Chew}, {Martin}, {Pollacco}  \& {Udry}}{{Armstrong}
  et~al.}{2014}]{armstrong2014}
{Armstrong} D.~J.,  {Osborn} H.~P.,  {Brown} D.~J.~A.,  {Faedi} F.,  {G{\'o}mez
  Maqueo Chew} Y.,  {Martin} D.~V.,  {Pollacco} D.,   {Udry} S.,  2014, \mn@doi
  [\mnras] {10.1093/mnras/stu1570}, \href
  {https://ui.adsabs.harvard.edu/abs/2014MNRAS.444.1873A} {444, 1873}

\bibitem[\protect\citeauthoryear{{Artymowicz} \& {Lubow}}{{Artymowicz} \&
  {Lubow}}{1994}]{Artymowicz1994}
{Artymowicz} P.,  {Lubow} S.~H.,  1994, \mn@doi [\apj] {10.1086/173679}, \href
  {https://ui.adsabs.harvard.edu/abs/1994ApJ...421..651A} {421, 651}

\bibitem[\protect\citeauthoryear{{Barbosa}, {Winter}, {Amarante}, {Izidoro},
  {Domingos}  \& {Macau}}{{Barbosa} et~al.}{2020}]{Barbosa2020}
{Barbosa} G.~O.,  {Winter} O.~C.,  {Amarante} A.,  {Izidoro} A.,  {Domingos}
  R.~C.,   {Macau} E.~E.~N.,  2020, \mn@doi [\mnras] {10.1093/mnras/staa757},
  \href {https://ui.adsabs.harvard.edu/abs/2020MNRAS.494.1045B} {494, 1045}

\bibitem[\protect\citeauthoryear{{Bashi}, {Helled}, {Zucker}  \&
  {Mordasini}}{{Bashi} et~al.}{2017}]{Bashi2017}
{Bashi} D.,  {Helled} R.,  {Zucker} S.,   {Mordasini} C.,  2017, \mn@doi [\aap]
  {10.1051/0004-6361/201629922}, \href
  {https://ui.adsabs.harvard.edu/abs/2017A&A...604A..83B} {604, A83}

\bibitem[\protect\citeauthoryear{{Batygin}}{{Batygin}}{2015}]{Batygin2015}
{Batygin} K.,  2015, \mn@doi [\mnras] {10.1093/mnras/stv1063}, \href
  {https://ui.adsabs.harvard.edu/abs/2015MNRAS.451.2589B} {451, 2589}

\bibitem[\protect\citeauthoryear{{Bennett} et~al.,}{{Bennett}
  et~al.}{2016}]{Bennett2016}
{Bennett} D.~P.,  et~al., 2016, \mn@doi [\aj] {10.3847/0004-6256/152/5/125},
  \href {https://ui.adsabs.harvard.edu/abs/2016AJ....152..125B} {152, 125}

\bibitem[\protect\citeauthoryear{{Bodman} \& {Quillen}}{{Bodman} \&
  {Quillen}}{2014}]{bodman2014}
{Bodman} E. H.~L.,  {Quillen} A.~C.,  2014, \mn@doi [\mnras]
  {10.1093/mnras/stu385}, \href
  {https://ui.adsabs.harvard.edu/abs/2014MNRAS.440.1753B} {440, 1753}

\bibitem[\protect\citeauthoryear{{Chatterjee}, {Ford}, {Matsumura}  \&
  {Rasio}}{{Chatterjee} et~al.}{2008}]{chatterjee2008}
{Chatterjee} S.,  {Ford} E.~B.,  {Matsumura} S.,   {Rasio} F.~A.,  2008,
  \mn@doi [\apj] {10.1086/590227}, \href
  {https://ui.adsabs.harvard.edu/abs/2008ApJ...686..580C} {686, 580}

\bibitem[\protect\citeauthoryear{{Chen}, {Franchini}, {Lubow}  \&
  {Martin}}{{Chen} et~al.}{2019}]{cheng2019}
{Chen} C.,  {Franchini} A.,  {Lubow} S.~H.,   {Martin} R.~G.,  2019, \mn@doi
  [\mnras] {10.1093/mnras/stz2948}, \href
  {https://ui.adsabs.harvard.edu/abs/2019MNRAS.490.5634C} {490, 5634}

\bibitem[\protect\citeauthoryear{{Chiang} \& {Laughlin}}{{Chiang} \&
  {Laughlin}}{2013}]{Chiang2013}
{Chiang} E.,  {Laughlin} G.,  2013, \mn@doi [\mnras] {10.1093/mnras/stt424},
  \href {https://ui.adsabs.harvard.edu/abs/2013MNRAS.431.3444C} {431, 3444}

\bibitem[\protect\citeauthoryear{{Childs} \& {Martin}}{{Childs} \&
  {Martin}}{2021a}]{Childs2021b}
{Childs} A.,  {Martin} R.,  2021a, arXiv e-prints, \href
  {https://ui.adsabs.harvard.edu/abs/2021arXiv210911653C} {p. arXiv:2109.11653}

\bibitem[\protect\citeauthoryear{{Childs} \& {Martin}}{{Childs} \&
  {Martin}}{2021b}]{Childs2021a}
{Childs} A.~C.,  {Martin} R.~G.,  2021b, \mn@doi [\mnras]
  {10.1093/mnras/stab2419}, \href
  {https://ui.adsabs.harvard.edu/abs/2021MNRAS.507.3461C} {507, 3461}

\bibitem[\protect\citeauthoryear{{Doolin} \& {Blundell}}{{Doolin} \&
  {Blundell}}{2011}]{Doolin2011}
{Doolin} S.,  {Blundell} K.~M.,  2011, \mn@doi [\mnras]
  {10.1111/j.1365-2966.2011.19657.x}, \href
  {https://ui.adsabs.harvard.edu/abs/2011MNRAS.418.2656D} {418, 2656}

\bibitem[\protect\citeauthoryear{{Doyle} et~al.,}{{Doyle}
  et~al.}{2011}]{doyle2011}
{Doyle} L.~R.,  et~al., 2011, \mn@doi [Science] {10.1126/science.1210923},
  \href {https://ui.adsabs.harvard.edu/abs/2011Sci...333.1602D} {333, 1602}

\bibitem[\protect\citeauthoryear{Duncan, Quinn  \& Tremaine}{Duncan
  et~al.}{1989}]{DUNCAN1989}
Duncan M.,  Quinn T.,   Tremaine S.,  1989, \mn@doi [Icarus]
  {https://doi.org/10.1016/0019-1035(89)90047-X}, 82, 402

\bibitem[\protect\citeauthoryear{{Dvorak}}{{Dvorak}}{1984}]{dvorak1984}
{Dvorak} R.,  1984, \mn@doi [Celestial Mechanics] {10.1007/BF01235815}, \href
  {https://ui.adsabs.harvard.edu/abs/1984CeMec..34..369D} {34, 369}

\bibitem[\protect\citeauthoryear{{Farago} \& {Laskar}}{{Farago} \&
  {Laskar}}{2010}]{farago10}
{Farago} F.,  {Laskar} J.,  2010, \mn@doi [\mnras]
  {10.1111/j.1365-2966.2009.15711.x}, \href
  {https://ui.adsabs.harvard.edu/abs/2010MNRAS.401.1189F} {401, 1189}

\bibitem[\protect\citeauthoryear{{Fleming}, {Barnes}, {Graham}, {Luger}  \&
  {Quinn}}{{Fleming} et~al.}{2018}]{fleming2018}
{Fleming} D.~P.,  {Barnes} R.,  {Graham} D.~E.,  {Luger} R.,   {Quinn} T.~R.,
  2018, \mn@doi [\apj] {10.3847/1538-4357/aabd38}, \href
  {https://ui.adsabs.harvard.edu/abs/2018ApJ...858...86F} {858, 86}

\bibitem[\protect\citeauthoryear{{Fulton} et~al.,}{{Fulton}
  et~al.}{2017}]{fulton2017}
{Fulton} B.~J.,  et~al., 2017, \mn@doi [\aj] {10.3847/1538-3881/aa80eb}, \href
  {https://ui.adsabs.harvard.edu/abs/2017AJ....154..109F} {154, 109}

\bibitem[\protect\citeauthoryear{{Goldreich} \& {Tremaine}}{{Goldreich} \&
  {Tremaine}}{1979}]{Goldreich1979}
{Goldreich} P.,  {Tremaine} S.,  1979, \mn@doi [\apj] {10.1086/157448}, \href
  {https://ui.adsabs.harvard.edu/abs/1979ApJ...233..857G} {233, 857}

\bibitem[\protect\citeauthoryear{{Hansen} \& {Murray}}{{Hansen} \&
  {Murray}}{2013}]{Hansen2013}
{Hansen} B. M.~S.,  {Murray} N.,  2013, \mn@doi [\apj]
  {10.1088/0004-637X/775/1/53}, \href
  {https://ui.adsabs.harvard.edu/abs/2013ApJ...775...53H} {775, 53}

\bibitem[\protect\citeauthoryear{{Hinse}, {Haghighipour}, {Kostov}  \&
  {Go{\'z}dziewski}}{{Hinse} et~al.}{2015}]{Hinse2015}
{Hinse} T.~C.,  {Haghighipour} N.,  {Kostov} V.~B.,   {Go{\'z}dziewski} K.,
  2015, \mn@doi [\apj] {10.1088/0004-637X/799/1/88}, \href
  {https://ui.adsabs.harvard.edu/abs/2015ApJ...799...88H} {799, 88}

\bibitem[\protect\citeauthoryear{{Holman} \& {Wiegert}}{{Holman} \&
  {Wiegert}}{1999}]{holman1999}
{Holman} M.~J.,  {Wiegert} P.~A.,  1999, \mn@doi [\aj] {10.1086/300695}, \href
  {https://ui.adsabs.harvard.edu/abs/1999AJ....117..621H} {117, 621}

\bibitem[\protect\citeauthoryear{{Huang} \& {Ormel}}{{Huang} \&
  {Ormel}}{2021}]{Huang2021}
{Huang} S.,  {Ormel} C.~W.,  2021, arXiv e-prints, \href
  {https://ui.adsabs.harvard.edu/abs/2021arXiv210910984H} {p. arXiv:2109.10984}

\bibitem[\protect\citeauthoryear{{Inamdar} \& {Schlichting}}{{Inamdar} \&
  {Schlichting}}{2015}]{Inamdar2015}
{Inamdar} N.~K.,  {Schlichting} H.~E.,  2015, \mn@doi [\mnras]
  {10.1093/mnras/stv030}, \href
  {https://ui.adsabs.harvard.edu/abs/2015MNRAS.448.1751I} {448, 1751}

\bibitem[\protect\citeauthoryear{{Izidoro}, {Bitsch}, {Raymond}, {Johansen},
  {Morbidelli}, {Lambrechts}  \& {Jacobson}}{{Izidoro}
  et~al.}{2021}]{Izidoro2021}
{Izidoro} A.,  {Bitsch} B.,  {Raymond} S.~N.,  {Johansen} A.,  {Morbidelli} A.,
   {Lambrechts} M.,   {Jacobson} S.~A.,  2021, \mn@doi [\aap]
  {10.1051/0004-6361/201935336}, \href
  {https://ui.adsabs.harvard.edu/abs/2021A&A...650A.152I} {650, A152}

\bibitem[\protect\citeauthoryear{Kasdin}{Kasdin}{1995}]{Kasdin1995}
Kasdin N.,  1995, \mn@doi [Proceedings of the IEEE] {10.1109/5.381848}, 83, 802

\bibitem[\protect\citeauthoryear{{Ketchum}, {Adams}  \& {Bloch}}{{Ketchum}
  et~al.}{2011}]{Ketchum2011}
{Ketchum} J.~A.,  {Adams} F.~C.,   {Bloch} A.~M.,  2011, \mn@doi [\apj]
  {10.1088/0004-637X/726/1/53}, \href
  {https://ui.adsabs.harvard.edu/abs/2011ApJ...726...53K} {726, 53}

\bibitem[\protect\citeauthoryear{{Kley} \& {Haghighipour}}{{Kley} \&
  {Haghighipour}}{2015}]{Kley2015}
{Kley} W.,  {Haghighipour} N.,  2015, \mn@doi [\aap]
  {10.1051/0004-6361/201526648}, \href
  {https://ui.adsabs.harvard.edu/abs/2015A&A...581A..20K} {581, A20}

\bibitem[\protect\citeauthoryear{{Kley}, {Peitz}  \& {Bryden}}{{Kley}
  et~al.}{2004}]{Kley2004}
{Kley} W.,  {Peitz} J.,   {Bryden} G.,  2004, \mn@doi [\aap]
  {10.1051/0004-6361:20031589}, \href
  {https://ui.adsabs.harvard.edu/abs/2004A&A...414..735K} {414, 735}

\bibitem[\protect\citeauthoryear{{Kostov}, {McCullough}, {Hinse}, {Tsvetanov},
  {H{\'e}brard}, {D{\'\i}az}, {Deleuil}  \& {Valenti}}{{Kostov}
  et~al.}{2013}]{kostov2013}
{Kostov} V.~B.,  {McCullough} P.~R.,  {Hinse} T.~C.,  {Tsvetanov} Z.~I.,
  {H{\'e}brard} G.,  {D{\'\i}az} R.~F.,  {Deleuil} M.,   {Valenti} J.~A.,
  2013, \mn@doi [\apj] {10.1088/0004-637X/770/1/52}, \href
  {https://ui.adsabs.harvard.edu/abs/2013ApJ...770...52K} {770, 52}

\bibitem[\protect\citeauthoryear{{Kostov} et~al.,}{{Kostov}
  et~al.}{2014}]{kostov2014}
{Kostov} V.~B.,  et~al., 2014, \mn@doi [\apj] {10.1088/0004-637X/784/1/14},
  \href {https://ui.adsabs.harvard.edu/abs/2014ApJ...784...14K} {784, 14}

\bibitem[\protect\citeauthoryear{{Kostov} et~al.,}{{Kostov}
  et~al.}{2016}]{kostov2016}
{Kostov} V.~B.,  et~al., 2016, \mn@doi [\apj] {10.3847/0004-637X/827/1/86},
  \href {https://ui.adsabs.harvard.edu/abs/2016ApJ...827...86K} {827, 86}

\bibitem[\protect\citeauthoryear{{Kostov} et~al.,}{{Kostov}
  et~al.}{2020a}]{kostov2020b}
{Kostov} V.~B.,  et~al., 2020a, arXiv e-prints, \href
  {https://ui.adsabs.harvard.edu/abs/2020arXiv200802756K} {p. arXiv:2008.02756}

\bibitem[\protect\citeauthoryear{Kostov et~al.,}{Kostov
  et~al.}{2020b}]{kostov2020}
Kostov V.~B.,  et~al., 2020b, arXiv e-prints, 2004, arXiv:2004.07783

\bibitem[\protect\citeauthoryear{{Kostov} et~al.,}{{Kostov}
  et~al.}{2021}]{kostov2020c}
{Kostov} V.~B.,  et~al., 2021, arXiv e-prints, \href
  {https://ui.adsabs.harvard.edu/abs/2021arXiv210508614K} {p. arXiv:2105.08614}

\bibitem[\protect\citeauthoryear{{Lam} \& {Kipping}}{{Lam} \&
  {Kipping}}{2018}]{Lam2018}
{Lam} C.,  {Kipping} D.,  2018, \mn@doi [\mnras] {10.1093/mnras/sty022}, \href
  {https://ui.adsabs.harvard.edu/abs/2018MNRAS.476.5692L} {476, 5692}

\bibitem[\protect\citeauthoryear{{Lee} \& {Chiang}}{{Lee} \&
  {Chiang}}{2016}]{Lee2016}
{Lee} E.~J.,  {Chiang} E.,  2016, \mn@doi [\apj] {10.3847/0004-637X/817/2/90},
  \href {https://ui.adsabs.harvard.edu/abs/2016ApJ...817...90L} {817, 90}

\bibitem[\protect\citeauthoryear{{Lee} \& {Peale}}{{Lee} \&
  {Peale}}{2002}]{lee2002}
{Lee} M.~H.,  {Peale} S.~J.,  2002, \mn@doi [\apj] {10.1086/338504}, \href
  {https://ui.adsabs.harvard.edu/abs/2002ApJ...567..596L} {567, 596}

\bibitem[\protect\citeauthoryear{Leung \& Lee}{Leung \& Lee}{2013}]{leung2013}
Leung G. C.~K.,  Lee M.~H.,  2013, \mn@doi [The Astrophysical Journal]
  {10.1088/0004-637x/763/2/107}, 763, 107

\bibitem[\protect\citeauthoryear{{Li}, {Holman}  \& {Tao}}{{Li}
  et~al.}{2016}]{li2016}
{Li} G.,  {Holman} M.~J.,   {Tao} M.,  2016, \mn@doi [\apj]
  {10.3847/0004-637X/831/1/96}, \href
  {https://ui.adsabs.harvard.edu/abs/2016ApJ...831...96L} {831, 96}

\bibitem[\protect\citeauthoryear{{Lin}, {Bodenheimer}  \& {Richardson}}{{Lin}
  et~al.}{1996}]{lin1996}
{Lin} D.~N.~C.,  {Bodenheimer} P.,   {Richardson} D.~C.,  1996, \mn@doi [\nat]
  {10.1038/380606a0}, \href
  {https://ui.adsabs.harvard.edu/abs/1996Natur.380..606L} {380, 606}

\bibitem[\protect\citeauthoryear{{Lines}, {Leinhardt}, {Paardekooper},
  {Baruteau}  \& {Thebault}}{{Lines} et~al.}{2014}]{Lines2014}
{Lines} S.,  {Leinhardt} Z.~M.,  {Paardekooper} S.,  {Baruteau} C.,
  {Thebault} P.,  2014, \mn@doi [\apjl] {10.1088/2041-8205/782/1/L11}, \href
  {https://ui.adsabs.harvard.edu/abs/2014ApJ...782L..11L} {782, L11}

\bibitem[\protect\citeauthoryear{{Lubow} \& {Ida}}{{Lubow} \&
  {Ida}}{2010}]{lubow2010}
{Lubow} S.~H.,  {Ida} S.,  2010, arXiv e-prints, \href
  {https://ui.adsabs.harvard.edu/abs/2010arXiv1004.4137L} {p. arXiv:1004.4137}

\bibitem[\protect\citeauthoryear{{Luhn}, {Penny}  \& {Gaudi}}{{Luhn}
  et~al.}{2016}]{Luhn2016}
{Luhn} J.~K.,  {Penny} M.~T.,   {Gaudi} B.~S.,  2016, \mn@doi [\apj]
  {10.3847/0004-637X/827/1/61}, \href
  {https://ui.adsabs.harvard.edu/abs/2016ApJ...827...61L} {827, 61}

\bibitem[\protect\citeauthoryear{{MacDonald}, {Dawson}, {Morrison}, {Lee}  \&
  {Khandelwal}}{{MacDonald} et~al.}{2020}]{MacDonald2020}
{MacDonald} M.~G.,  {Dawson} R.~I.,  {Morrison} S.~J.,  {Lee} E.~J.,
  {Khandelwal} A.,  2020, \mn@doi [\apj] {10.3847/1538-4357/ab6f04}, \href
  {https://ui.adsabs.harvard.edu/abs/2020ApJ...891...20M} {891, 20}

\bibitem[\protect\citeauthoryear{{MacDonald}, {Shakespeare}  \&
  {Ragozzine}}{{MacDonald} et~al.}{2021}]{MacDonald2021}
{MacDonald} M.~G.,  {Shakespeare} C.~J.,   {Ragozzine} D.,  2021, \mn@doi [\aj]
  {10.3847/1538-3881/ac12d5}, \href
  {https://ui.adsabs.harvard.edu/abs/2021AJ....162..114M} {162, 114}

\bibitem[\protect\citeauthoryear{{Malhotra}}{{Malhotra}}{1998}]{Malhotra1998}
{Malhotra} R.,  1998, in {Lazzaro} D.,  {Vieira Martins} R.,  {Ferraz-Mello}
  S.,   {Fernandez} J.,  eds,  Astronomical Society of the Pacific Conference
  Series Vol. 149, Solar System Formation and Evolution. p.~37

\bibitem[\protect\citeauthoryear{{Mardling}}{{Mardling}}{2008}]{Mardling2008}
{Mardling} R.~A.,  2008, {Resonance, Chaos and Stability: The Three-Body
  Problem in Astrophysics}.
p.~59, \mn@doi{10.1007/978-1-4020-8431-7\_3}

\bibitem[\protect\citeauthoryear{Mardling}{Mardling}{2013}]{mardling2013}
Mardling R.~A.,  2013, \mn@doi [Monthly Notices of the Royal Astronomical
  Society] {10.1093/mnras/stt1438}, 435, 2187–2226

\bibitem[\protect\citeauthoryear{{Martin}}{{Martin}}{2017}]{martin2017}
{Martin} D.~V.,  2017, \mn@doi [\mnras] {10.1093/mnras/stw2851}, \href
  {https://ui.adsabs.harvard.edu/abs/2017MNRAS.465.3235M} {465, 3235}

\bibitem[\protect\citeauthoryear{Martin}{Martin}{2018}]{martin2018}
Martin D.~V.,  2018, Populations of Planets in Multiple Star Systems.
Springer International Publishing, Cham, pp 2035--2060,
  \mn@doi{10.1007/978-3-319-55333-7_156}, \url
  {https://doi.org/10.1007/978-3-319-55333-7_156}

\bibitem[\protect\citeauthoryear{Martin \& Fabrycky}{Martin \&
  Fabrycky}{2021}]{martin2021}
Martin D.~V.,  Fabrycky D.~C.,  2021, Searching for Small Circumbinary Planets
  I. The STANLEY Automated Algorithm and No New Planets in Existing Systems
  (\mn@eprint {arXiv} {2101.03186})

\bibitem[\protect\citeauthoryear{{Martin} \& {Triaud}}{{Martin} \&
  {Triaud}}{2014}]{martin2014}
{Martin} D.~V.,  {Triaud} A. H.~M.~J.,  2014, \mn@doi [\aap]
  {10.1051/0004-6361/201323112}, \href
  {https://ui.adsabs.harvard.edu/abs/2014A&A...570A..91M} {570, A91}

\bibitem[\protect\citeauthoryear{{Martin} \& {Triaud}}{{Martin} \&
  {Triaud}}{2016}]{martin2016}
{Martin} D.~V.,  {Triaud} A. H.~M.~J.,  2016, \mn@doi [\mnras]
  {10.1093/mnrasl/slv139}, \href
  {https://ui.adsabs.harvard.edu/abs/2016MNRAS.455L..46M} {455, L46}

\bibitem[\protect\citeauthoryear{Martin et~al.,}{Martin
  et~al.}{2019}]{martin2019}
Martin D.~V.,  et~al., 2019, \mn@doi [Astronomy and Astrophysics]
  {10.1051/0004-6361/201833669}, 624, A68

\bibitem[\protect\citeauthoryear{Meschiari}{Meschiari}{2014}]{Meschiari2014}
Meschiari S.,  2014, \mn@doi [The Astrophysical Journal]
  {10.1088/0004-637x/790/1/41}, 790, 41

\bibitem[\protect\citeauthoryear{{Miranda} \& {Lai}}{{Miranda} \&
  {Lai}}{2015}]{miranda2015}
{Miranda} R.,  {Lai} D.,  2015, \mn@doi [\mnras] {10.1093/mnras/stv1450}, \href
  {https://ui.adsabs.harvard.edu/abs/2015MNRAS.452.2396M} {452, 2396}

\bibitem[\protect\citeauthoryear{{Mudryk} \& {Wu}}{{Mudryk} \&
  {Wu}}{2006}]{Mudryk2006}
{Mudryk} L.~R.,  {Wu} Y.,  2006, \mn@doi [\apj] {10.1086/499347}, \href
  {https://ui.adsabs.harvard.edu/abs/2006ApJ...639..423M} {639, 423}

\bibitem[\protect\citeauthoryear{{Murray} \& {Dermott}}{{Murray} \&
  {Dermott}}{1999}]{murray1999}
{Murray} C.~D.,  {Dermott} S.~F.,  1999, {Solar system dynamics}

\bibitem[\protect\citeauthoryear{{Mustill} \& {Wyatt}}{{Mustill} \&
  {Wyatt}}{2011}]{Mustill2016}
{Mustill} A.~J.,  {Wyatt} M.~C.,  2011, \mn@doi [\mnras]
  {10.1111/j.1365-2966.2011.18201.x}, \href
  {https://ui.adsabs.harvard.edu/abs/2011MNRAS.413..554M} {413, 554}

\bibitem[\protect\citeauthoryear{{Nelson}}{{Nelson}}{2018}]{Nelson2018}
{Nelson} R.~P.,  2018, {Planetary Migration in Protoplanetary Disks}.
p.~139, \mn@doi{10.1007/978-3-319-55333-7_139}

\bibitem[\protect\citeauthoryear{{Ogihara}, {Morbidelli}  \&
  {Guillot}}{{Ogihara} et~al.}{2015}]{Ogihara2015}
{Ogihara} M.,  {Morbidelli} A.,   {Guillot} T.,  2015, \mn@doi [\aap]
  {10.1051/0004-6361/201525884}, \href
  {https://ui.adsabs.harvard.edu/abs/2015A&A...578A..36O} {578, A36}

\bibitem[\protect\citeauthoryear{Orosz et~al.,}{Orosz
  et~al.}{2012a}]{orosz2012b}
Orosz J.~A.,  et~al., 2012a, \mn@doi [Science] {10.1126/science.1228380}, 337,
  1511

\bibitem[\protect\citeauthoryear{{Orosz} et~al.,}{{Orosz}
  et~al.}{2012b}]{orosz2012}
{Orosz} J.~A.,  et~al., 2012b, \mn@doi [\apj] {10.1088/0004-637X/758/2/87},
  \href {https://ui.adsabs.harvard.edu/abs/2012ApJ...758...87O} {758, 87}

\bibitem[\protect\citeauthoryear{{Orosz} et~al.,}{{Orosz}
  et~al.}{2019}]{orosz2019}
{Orosz} J.~A.,  et~al., 2019, \mn@doi [\aj] {10.3847/1538-3881/ab0ca0}, \href
  {https://ui.adsabs.harvard.edu/abs/2019AJ....157..174O} {157, 174}

\bibitem[\protect\citeauthoryear{{Paardekooper}, {Baruteau}, {Crida}  \&
  {Kley}}{{Paardekooper} et~al.}{2010}]{Paardekooper2010}
{Paardekooper} S.~J.,  {Baruteau} C.,  {Crida} A.,   {Kley} W.,  2010, \mn@doi
  [\mnras] {10.1111/j.1365-2966.2009.15782.x}, \href
  {https://ui.adsabs.harvard.edu/abs/2010MNRAS.401.1950P} {401, 1950}

\bibitem[\protect\citeauthoryear{{Paardekooper}, {Baruteau}  \&
  {Kley}}{{Paardekooper} et~al.}{2011}]{Paardekooper2011}
{Paardekooper} S.~J.,  {Baruteau} C.,   {Kley} W.,  2011, \mn@doi [\mnras]
  {10.1111/j.1365-2966.2010.17442.x}, \href
  {https://ui.adsabs.harvard.edu/abs/2011MNRAS.410..293P} {410, 293}

\bibitem[\protect\citeauthoryear{Paardekooper, Leinhardt, Th{\'{e}}bault  \&
  Baruteau}{Paardekooper et~al.}{2012}]{Paardekooper2012}
Paardekooper S.-J.,  Leinhardt Z.~M.,  Th{\'{e}}bault P.,   Baruteau C.,  2012,
  \mn@doi [The Astrophysical Journal] {10.1088/2041-8205/754/1/l16}, 754, L16

\bibitem[\protect\citeauthoryear{Papaloizou \& Larwood}{Papaloizou \&
  Larwood}{2000}]{Papaloizou2000}
Papaloizou J. C.~B.,  Larwood J.~D.,  2000, \mn@doi [Monthly Notices of the
  Royal Astronomical Society] {10.1046/j.1365-8711.2000.03466.x}, 315,
  823–833

\bibitem[\protect\citeauthoryear{Penzlin, Kley  \& Nelson}{Penzlin
  et~al.}{2021}]{Penzlin2021}
Penzlin A. B.~T.,  Kley W.,   Nelson R.~P.,  2021, \mn@doi [Astronomy &
  Astrophysics] {10.1051/0004-6361/202039319}, 645, A68

\bibitem[\protect\citeauthoryear{{Petigura}, {Howard}  \& {Marcy}}{{Petigura}
  et~al.}{2013}]{petigura2013}
{Petigura} E.~A.,  {Howard} A.~W.,   {Marcy} G.~W.,  2013, \mn@doi [Proceedings
  of the National Academy of Science] {10.1073/pnas.1319909110}, \href
  {https://ui.adsabs.harvard.edu/abs/2013PNAS..11019273P} {110, 19273}

\bibitem[\protect\citeauthoryear{Pierens \& Nelson}{Pierens \&
  Nelson}{2008}]{Pierens2008}
Pierens A.,  Nelson R.~P.,  2008, \mn@doi [Astronomy & Astrophysics]
  {10.1051/0004-6361:200809453}, 483, 633–642

\bibitem[\protect\citeauthoryear{Pierens \& Nelson}{Pierens \&
  Nelson}{2013}]{pierens2013}
Pierens A.,  Nelson R.,  2013, \mn@doi [Astronomy and Astrophysics]
  {10.1051/0004-6361/201321777}, 556

\bibitem[\protect\citeauthoryear{Pierens, McNally  \& Nelson}{Pierens
  et~al.}{2020}]{pierens2020}
Pierens A.,  McNally C.~P.,   Nelson R.~P.,  2020, \mn@doi [Monthly Notices of
  the Royal Astronomical Society] {10.1093/mnras/staa1550}, 496, 2849

\bibitem[\protect\citeauthoryear{{Pierens}, {Nelson}  \& {McNally}}{{Pierens}
  et~al.}{2021}]{Pierens2021}
{Pierens} A.,  {Nelson} R.~P.,   {McNally} C.~P.,  2021, arXiv e-prints, \href
  {https://ui.adsabs.harvard.edu/abs/2021arXiv210914466P} {p. arXiv:2109.14466}

\bibitem[\protect\citeauthoryear{{Pilat-Lohinger}, {Funk}  \&
  {Dvorak}}{{Pilat-Lohinger} et~al.}{2003}]{Pilat-Lohinge2003}
{Pilat-Lohinger} E.,  {Funk} B.,   {Dvorak} R.,  2003, \mn@doi [\aap]
  {10.1051/0004-6361:20021811}, \href
  {https://ui.adsabs.harvard.edu/abs/2003A&A...400.1085P} {400, 1085}

\bibitem[\protect\citeauthoryear{{Quarles}, {Satyal}, {Kostov}, {Kaib}  \&
  {Haghighipour}}{{Quarles} et~al.}{2018}]{quarles2018}
{Quarles} B.,  {Satyal} S.,  {Kostov} V.,  {Kaib} N.,   {Haghighipour} N.,
  2018, \mn@doi [\apj] {10.3847/1538-4357/aab264}, \href
  {https://ui.adsabs.harvard.edu/abs/2018ApJ...856..150Q} {856, 150}

\bibitem[\protect\citeauthoryear{{Ragusa}, {Alexander}, {Calcino}, {Hirsh}  \&
  {Price}}{{Ragusa} et~al.}{2020}]{ragusa2020}
{Ragusa} E.,  {Alexander} R.,  {Calcino} J.,  {Hirsh} K.,   {Price} D.~J.,
  2020, \mn@doi [\mnras] {10.1093/mnras/staa2954}, \href
  {https://ui.adsabs.harvard.edu/abs/2020MNRAS.499.3362R} {499, 3362}

\bibitem[\protect\citeauthoryear{{Raymond}, {Boulet}, {Izidoro}, {Esteves}  \&
  {Bitsch}}{{Raymond} et~al.}{2018}]{Raymond2018}
{Raymond} S.~N.,  {Boulet} T.,  {Izidoro} A.,  {Esteves} L.,   {Bitsch} B.,
  2018, \mn@doi [\mnras] {10.1093/mnrasl/sly100}, \href
  {https://ui.adsabs.harvard.edu/abs/2018MNRAS.479L..81R} {479, L81}

\bibitem[\protect\citeauthoryear{{Rein}}{{Rein}}{2010}]{Rein2010PhD}
{Rein} H.,  2010, PhD thesis, -

\bibitem[\protect\citeauthoryear{{Rein}}{{Rein}}{2012}]{rein2012PERIODRATIOS}
{Rein} H.,  2012, \mn@doi [\mnras] {10.1111/j.1745-3933.2012.01337.x}, \href
  {https://ui.adsabs.harvard.edu/abs/2012MNRAS.427L..21R} {427, L21}

\bibitem[\protect\citeauthoryear{{Rein} \& {Liu}}{{Rein} \&
  {Liu}}{2012}]{rein2012REBOUND}
{Rein} H.,  {Liu} S.~F.,  2012, \mn@doi [\aap] {10.1051/0004-6361/201118085},
  \href {https://ui.adsabs.harvard.edu/abs/2012A&A...537A.128R} {537, A128}

\bibitem[\protect\citeauthoryear{{Rein} \& {Papaloizou}}{{Rein} \&
  {Papaloizou}}{2009}]{Rein2009}
{Rein} H.,  {Papaloizou} J.~C.~B.,  2009, \mn@doi [\aap]
  {10.1051/0004-6361/200811330}, \href
  {https://ui.adsabs.harvard.edu/abs/2009A&A...497..595R} {497, 595}

\bibitem[\protect\citeauthoryear{{Rein} \& {Spiegel}}{{Rein} \&
  {Spiegel}}{2015}]{rein2015IAS15}
{Rein} H.,  {Spiegel} D.~S.,  2015, \mn@doi [\mnras] {10.1093/mnras/stu2164},
  \href {https://ui.adsabs.harvard.edu/abs/2015MNRAS.446.1424R} {446, 1424}

\bibitem[\protect\citeauthoryear{{Rein} \& {Tamayo}}{{Rein} \&
  {Tamayo}}{2015}]{rein2015WHFAST}
{Rein} H.,  {Tamayo} D.,  2015, \mn@doi [\mnras] {10.1093/mnras/stv1257}, \href
  {https://ui.adsabs.harvard.edu/abs/2015MNRAS.452..376R} {452, 376}

\bibitem[\protect\citeauthoryear{{Rein}, {Payne}, {Veras}  \& {Ford}}{{Rein}
  et~al.}{2012}]{rein2012FOURTHREE}
{Rein} H.,  {Payne} M.~J.,  {Veras} D.,   {Ford} E.~B.,  2012, \mn@doi [\mnras]
  {10.1111/j.1365-2966.2012.21798.x}, \href
  {https://ui.adsabs.harvard.edu/abs/2012MNRAS.426..187R} {426, 187}

\bibitem[\protect\citeauthoryear{{Schwamb} et~al.,}{{Schwamb}
  et~al.}{2013}]{schwamb2013}
{Schwamb} M.~E.,  et~al., 2013, \mn@doi [\apj] {10.1088/0004-637X/768/2/127},
  \href {https://ui.adsabs.harvard.edu/abs/2013ApJ...768..127S} {768, 127}

\bibitem[\protect\citeauthoryear{{Seager}, {Kuchner}, {Hier-Majumder}  \&
  {Militzer}}{{Seager} et~al.}{2007}]{Seager2007}
{Seager} S.,  {Kuchner} M.,  {Hier-Majumder} C.~A.,   {Militzer} B.,  2007,
  \mn@doi [\apj] {10.1086/521346}, \href
  {https://ui.adsabs.harvard.edu/abs/2007ApJ...669.1279S} {669, 1279}

\bibitem[\protect\citeauthoryear{{Shakura} \& {Sunyaev}}{{Shakura} \&
  {Sunyaev}}{1973}]{shakura1973}
{Shakura} N.~I.,  {Sunyaev} R.~A.,  1973, \aap, \href
  {https://ui.adsabs.harvard.edu/abs/1973A&A....24..337S} {500, 33}

\bibitem[\protect\citeauthoryear{{Socia} et~al.,}{{Socia}
  et~al.}{2020}]{socia2020}
{Socia} Q.~J.,  et~al., 2020, \mn@doi [\aj] {10.3847/1538-3881/ab665b}, \href
  {https://ui.adsabs.harvard.edu/abs/2020AJ....159...94S} {159, 94}

\bibitem[\protect\citeauthoryear{Sutherland \& Kratter}{Sutherland \&
  Kratter}{2019}]{sutherland2019}
Sutherland A.~P.,  Kratter K.~M.,  2019, \mn@doi [Monthly Notices of the Royal
  Astronomical Society] {10.1093/mnras/stz1503}, 487, 3288

\bibitem[\protect\citeauthoryear{{Tamayo}, {Rein}, {Shi}  \&
  {Hernandez}}{{Tamayo} et~al.}{2020}]{Tamayo2020}
{Tamayo} D.,  {Rein} H.,  {Shi} P.,   {Hernandez} D.~M.,  2020, \mn@doi
  [\mnras] {10.1093/mnras/stz2870}, \href
  {https://ui.adsabs.harvard.edu/abs/2020MNRAS.491.2885T} {491, 2885}

\bibitem[\protect\citeauthoryear{{Tanaka}, {Takeuchi}  \& {Ward}}{{Tanaka}
  et~al.}{2002}]{tanaka2002}
{Tanaka} H.,  {Takeuchi} T.,   {Ward} W.~R.,  2002, \mn@doi [\apj]
  {10.1086/324713}, \href
  {https://ui.adsabs.harvard.edu/abs/2002ApJ...565.1257T} {565, 1257}

\bibitem[\protect\citeauthoryear{{Terquem} \& {Papaloizou}}{{Terquem} \&
  {Papaloizou}}{2007}]{Terquem2007}
{Terquem} C.,  {Papaloizou} J. C.~B.,  2007, \mn@doi [\apj] {10.1086/509497},
  \href {https://ui.adsabs.harvard.edu/abs/2007ApJ...654.1110T} {654, 1110}

\bibitem[\protect\citeauthoryear{{Teyssandier} \& {Libert}}{{Teyssandier} \&
  {Libert}}{2020}]{Teyssandier2020}
{Teyssandier} J.,  {Libert} A.-S.,  2020, \mn@doi [\aap]
  {10.1051/0004-6361/202039038}, \href
  {https://ui.adsabs.harvard.edu/abs/2020A&A...643A..11T} {643, A11}

\bibitem[\protect\citeauthoryear{{Thun} \& {Kley}}{{Thun} \&
  {Kley}}{2018}]{thun2018}
{Thun} D.,  {Kley} W.,  2018, \mn@doi [\aap] {10.1051/0004-6361/201832804},
  \href {https://ui.adsabs.harvard.edu/abs/2018A&A...616A..47T} {616, A47}

\bibitem[\protect\citeauthoryear{{Ward}}{{Ward}}{1986}]{Ward1986}
{Ward} W.~R.,  1986, \mn@doi [\icarus] {10.1016/0019-1035(86)90182-X}, \href
  {https://ui.adsabs.harvard.edu/abs/1986Icar...67..164W} {67, 164}

\bibitem[\protect\citeauthoryear{{Ward}}{{Ward}}{1997}]{ward1997}
{Ward} W.~R.,  1997, \mn@doi [\icarus] {10.1006/icar.1996.5647}, \href
  {https://ui.adsabs.harvard.edu/abs/1997Icar..126..261W} {126, 261}

\bibitem[\protect\citeauthoryear{Welsh \& Orosz}{Welsh \&
  Orosz}{2018}]{welsh2018}
Welsh W.~F.,  Orosz J.~A.,  2018, Two Suns in the Sky: The Kepler Circumbinary
  Planets.
Springer International Publishing, Cham, pp 2749--2768,
  \mn@doi{10.1007/978-3-319-55333-7_34}, \url
  {https://doi.org/10.1007/978-3-319-55333-7_34}

\bibitem[\protect\citeauthoryear{{Welsh} et~al.,}{{Welsh}
  et~al.}{2012}]{welsh2012}
{Welsh} W.~F.,  et~al., 2012, \mn@doi [\nat] {10.1038/nature10768}, \href
  {https://ui.adsabs.harvard.edu/abs/2012Natur.481..475W} {481, 475}

\bibitem[\protect\citeauthoryear{{Welsh} et~al.,}{{Welsh}
  et~al.}{2015}]{welsh2015}
{Welsh} W.~F.,  et~al., 2015, \mn@doi [\apj] {10.1088/0004-637X/809/1/26},
  \href {https://ui.adsabs.harvard.edu/abs/2015ApJ...809...26W} {809, 26}

\bibitem[\protect\citeauthoryear{{Windemuth}, {Agol}, {Carter}, {Ford},
  {Haghighipour}, {Orosz}  \& {Welsh}}{{Windemuth}
  et~al.}{2019}]{windemuth2019b}
{Windemuth} D.,  {Agol} E.,  {Carter} J.,  {Ford} E.~B.,  {Haghighipour} N.,
  {Orosz} J.~A.,   {Welsh} W.~F.,  2019, \mn@doi [\mnras]
  {10.1093/mnras/stz2637}, \href
  {https://ui.adsabs.harvard.edu/abs/2019MNRAS.490.1313W} {490, 1313}

\makeatother
\end{thebibliography}

\bsp	
\label{lastpage}
\end{document}